\documentstyle[11pt,aaspp4]{article}
\begin{document}
%
\def\IUE{{\it IUE}}
\def\HST{{\it HST}}
\def\ASCA{{\it ASCA}}
\def\RXTE{{\it RXTE}}
\def\deg{$^{\rm o}$}
\def\degC{$^{\rm o}$C}
\def\arcsec{$''$}
\def\arcmin{$'$}
\def\arcsecpoint{\ifmmode ''\!. \else $''\!.$\fi}
\def\arcminpoint{\ifmmode '\!. \else $'\!.$\fi}
\def\kms{\ifmmode {\rm km\,s}^{-1} \else km\,s$^{-1}$\fi}
\def\cc{\ifmmode {\rm cm}^{-3} \else cm$^{-3}$\fi}
\def\Msun{\ifmmode M_{\odot} \else $M_{\odot}$\fi}
\def\Lsun{\ifmmode L_{\odot} \else $L_{\odot}$\fi}
\def\qo{\ifmmode q_{\rm o} \else $q_{\rm o}$\fi}
\def\Ho{\ifmmode H_{\rm o} \else $H_{\rm o}$\fi}
\def\ho{\ifmmode h_{\rm o} \else $h_{\rm o}$\fi}
\def\ltsim{\raisebox{-.5ex}{$\;\stackrel{<}{\sim}\;$}}
\def\gtsim{\raisebox{-.5ex}{$\;\stackrel{>}{\sim}\;$}}
\def\vFWHM{\ifmmode v_{\mbox{\tiny FWHM}} \else
            $v_{\mbox{\tiny FWHM}}$\fi}
\def\CCF{\ifmmode F_{\it CCF} \else $F_{\it CCF}$\fi}
\def\ACF{\ifmmode F_{\it ACF} \else $F_{\it ACF}$\fi}
\def\Halpha{\ifmmode {\rm H}\alpha \else H$\alpha$\fi}
\def\Hbeta{\ifmmode {\rm H}\beta \else H$\beta$\fi}
\def\Hgamma{\ifmmode {\rm H}\gamma \else H$\gamma$\fi}
\def\Hdelta{\ifmmode {\rm H}\delta \else H$\delta$\fi}
\def\Lya{\ifmmode {\rm Ly}\alpha \else Ly$\alpha$\fi}
\def\Lyb{\ifmmode {\rm Ly}\beta \else Ly$\beta$\fi}
\def\hi{H\,{\sc i}}
\def\hii{H\,{\sc ii}}
\def\hei{He\,{\sc i}}
\def\heii{He\,{\sc ii}}
\def\ci{C\,{\sc i}}
\def\cii{C\,{\sc ii}}
\def\ciii{\ifmmode {\rm C}\,{\sc iii} \else C\,{\sc iii}\fi}
\def\civ{\ifmmode {\rm C}\,{\sc iv} \else C\,{\sc iv}\fi}
\def\ni{N\,{\sc i}}
\def\nii{[N\,{\sc ii}]}
\def\niii{N\,{\sc iii}}
\def\niv{N\,{\sc iv}]}
\def\nv{N\,{\sc v}}
\def\oi{[O\,{\sc i}]}
\def\oii{O\,{\sc ii}}
\def\oiii{[O\,{\sc iii}]}
\def\o5007{[O\,{\sc iii}]\,$\lambda5007$}
\def\oiv{O\,{\sc iv}}
\def\ov{O\,{\sc v}}
\def\ovi{O\,{\sc vi}}
\def\nev{Ne\,{\sc v}}
\def\mgi{Mg\,{\sc i}}
\def\mgii{Mg\,{\sc ii}}
\def\siIII{Si\,{\sc iii}}
\def\siIV{Si\,{\sc iv}}
\def\si{S\,{\sc i}}
\def\sii{S\,{\sc ii}}
\def\siii{S\,{\sc iii}}
\def\caii{Ca\,{\sc ii}}
\def\feii{Fe\,{\sc ii}}
\def\fevii{Fe\,{\sc vii}}
\def\fex{Fe\,{\sc x}}
\def\aliii{Al\,{\sc iii}}
\def\lax    {${_<\atop^{\sim}}$}
\def\gax    {${_>\atop^{\sim}}$}

\title{Multiwavelength Monitoring of the Narrow-Line Seyfert 1 Galaxy 
Akn~564. II. Ultraviolet Continuum and Emission-line Variability}

\author{
S.~Collier,\altaffilmark{1}
D.M.~Crenshaw,\altaffilmark{2}
B.M.~Peterson,\altaffilmark{1}
W.N.~Brandt,\altaffilmark{3}
J.~Clavel,\altaffilmark{4}
R.~Edelson,\altaffilmark{5,6}
I.M.~George,\altaffilmark{7,8}
K.~Horne,\altaffilmark{9,10}
G.A.~Kriss,\altaffilmark{11}
S.~Mathur,\altaffilmark{1}
H.~Netzer,\altaffilmark{12}
P.T.~O'Brien,\altaffilmark{6}
R.W.~Pogge,\altaffilmark{1}
K.A.~Pounds,\altaffilmark{6}
P.~Romano,\altaffilmark{1} 
O.~Shemmer,\altaffilmark{12}
T.J.~Turner,\altaffilmark{7,8} and
W.~Wamsteker\altaffilmark{4}
\altaffiltext{1}{Department of Astronomy, The Ohio State University, 140 
West
18th Avenue, Columbus, OH 43210.}
\altaffiltext{2}{Catholic University of America and Laboratory for 
Astronomy and Solar
Physics, NASA Goddard Space Flight Center, Code 681, Greenbelt, MD 
20771}
\altaffiltext{5}{Department of Astronomy \& Astrophysics, The Pennsylvania
State University,
525 Davey Laboratory, University Park, PA  16802}
\altaffiltext{4}{ESA, P.O. Box 50727, 28080 Madrid, 
Spain}
\altaffiltext{5}{Department of Physics and Astronomy, University of 
California, Los
Angeles, CA 90095-1562}
\altaffiltext{6}{Department of Physics and Astronomy, University of 
Leicester,
University Road, Leicester, LE1 7RH, UK}
\altaffiltext{7}{Joint Center for Astrophysics, Department of Physics, 
University of Maryland, Baltimore County, 1000 Hilltop Circle, Baltimore, MD 21250}
\altaffiltext{8}{Laboratory for High Energy Astrophysics, Code 660,
NASA/Goddard Space Flight Center, Greenbelt, MD 20771}
\altaffiltext{9}{School of Physics and Astronomy, University of St.\ 
Andrews, St.\ Andrews, KY16 9SS, UK}
\altaffiltext{10}{Department of Astronomy, University of Texas, Austin, TX
78704}
\altaffiltext{11}{Space Telescope Science Institute, 3700 San Martin 
Drive, Baltimore, MD
21218}
\altaffiltext{12}{School of Physics and Astronomy and the Wise 
Observatory, The Raymond
and Beverly Sackler Faculty of Exact Sciences, Tel Aviv University, Tel 
Aviv 69978,
Israel}
}

\begin{abstract}

We present results of an intensive two-month campaign of approximately
daily spectrophotometric monitoring of the narrow-line Seyfert 1 galaxy
Akn~564 with the {\em Hubble Space Telescope} ({\em HST}). The fractional
variability amplitude of the continuum variations between 1365--3000\,\AA\
is $\sim6$\%, about a factor 3 less than that found in typical Seyfert 1
galaxies over a similar period of time. However, large amplitude, short
time-scale flaring behavior is evident, with trough-to-peak flux changes
of about 18\% in approximately 3 days. We present evidence for
wavelength-dependent continuum time delays, with the variations at
3000\,\AA\ lagging behind those at 1365\,\AA\ by about 1 day. These delays
may be interpreted as evidence for a stratified continuum reprocessing
region, possibly an accretion-disk structure. The \Lya\,$\lambda 1216$
emission-line exhibits flux variations of about 1\% amplitude. These
variations lag those at 1365\AA\ by \lax 3 days, and combining this with
the line width yields a virial black-hole mass limit of \lax $8 \times
10^{6}\,\Msun$. We caution, the low amplitude \Lya\,$\lambda1216$
variations may indicate the bulk of the emission region is at larger
radii. This scenario affects the veracity of our black hole mass upper
limit in an uncertain manner due to the unknown nature of the gas velocity
field. Our mass estimate is thus unreliable, however, it is consistent
with the independent estimate $M \sim 1 \times 10^{7}\,\Msun$ of Pounds et
al. (2001), based on a fluctuation power spectrum analysis of X-ray
variability in Akn 564. The black-hole mass and 5100\,\AA\ luminosity of
Akn~564 are consistent with the hypothesis that, relative to Seyfert 1
galaxies, NLS1s have lower black hole masses and higher accretion rates.
Other strong emission lines, e.g., \civ\,$\lambda1549$ and
\heii\,$\lambda1640$, are constrained to vary with amplitudes of $<$ 5\%.
This low-level of emission-line variability is different from most Seyfert
1 galaxies, which characteristically display variations of $\sim10$\% on
similar time scales.

\end{abstract}

\section{Introduction}

Narrow-line Seyfert 1 (NLS1) galaxies were first classified on the basis
of their narrow permitted optical emission lines, with \Hbeta\
full-width-half-maximum (FWHM) \lax 2000\,\kms (Osterbrock \& Pogge 1985). 
They also
exhibit
distinctive X-ray properties (e.g., Puchnarewicz et al.\ 1992; Boller et
al.\ 1996; Brandt, Mathur, \& Elvis 1997; Turner 1999; Leighly 1999a, b). These
include a steep soft
excess with
photon index
$\Gamma_{{\rm soft}} > 3$ (photon flux $P_{E} \propto E^{-\Gamma}$), a steep
hard power law with $\Gamma_{{\rm 
hard}} > 2.5$, and rapid, short time-scale variations (e.g., $\sim$ 30\%
in 1500 sec). Their UV/optical properties place them at one extreme of the
Boroson \& Green (1992) primary eigenvector that has been identified in a
principal component analysis;  specifically, NLS1 classification
correlates with strong optical \feii\ and weak \o5007\ emission. NLS1s are
thus potentially useful in identifying the underlying physics that defines
the primary eigenvector.

A number of scenarios have been posited to explain NLS1 properties:
\begin{enumerate}
\item NLS1s may have larger broad-line region (BLR) sizes compared to
Seyfert 1 (S1)
galaxies. The steeper soft excesses imply higher ionizing fluxes than 
S1
galaxies with comparable luminosities. Under certain conditions this 
may
lead to larger BLR sizes and hence smaller permitted line widths, 
assuming
they reflect virialized motions about the putative black hole
(Guilbert, Fabian, \& McCray 1983; Wandel \&
Boller 1998).
\item NLS1s may be low-inclination $i$, nearly face-on systems (Osterbrock 
\& Pogge 1985). The line widths, attributable to orbital, virialized 
motions in a common plane are decreased by $\sin i$, and the strong soft 
X-ray
fluxes are reconcilable with accretion disk models (e.g., Madau 1988). 
However, this orientation scenario has problems explaining the low \oiii\
luminosities often observed (e.g., Boroson \& Green 1992).
\item NLS1s may have relatively low black hole masses with higher than 
normal
accretion rates, compared to S1 galaxies (Pounds, Done, \& Osborne 1995). The
smaller line widths are attributable to the reduced
gravitational potential in which they form, and higher accretion
rates result in luminosities that are large for their masses.
\end{enumerate}

Balmer (\Hbeta) BLR sizes and masses have been measured for about 40 AGNs,
including five NLS1s (Wandel, Peterson, \& Malkan 1999; Kaspi et al.\
2000;  Peterson et al.\ 2000; Shemmer et al. 2001). These results suggest
that NLS1s have BLR
sizes comparable to those of S1s with similar optical luminosities.
Moreover, NLS1
and S1 galaxies delineate a broad mass-luminosity relationship with the
former sources populating the low mass extremum of this relationship. This
is consistent with the hypothesis that NLS1s are undermassive black hole
systems with higher accretion rates and/or sources viewed nearly face-on.

To distinguish between various models requires systematic measurements of
the black hole masses $M$, mass accretion rates $\dot{M}$, broad-line
region sizes $R_{{\rm BLR}}$, and source inclinations $i$ of NLS1s and
other S1s. These key parameters may all be measured directly or inferred
through application of echo mapping (reverberation) techniques (Blandford
\& McKee 1982; Peterson 2001) that use the relative responses of continuum
and emission-line components to constrain strongly the nature of the
responding emission regions. The responsivity-weighted BLR size is given
approximately by $R_{{\rm BLR}} = \tau c$, with $\tau$ the time lag,
measured from cross-correlation analysis, between the continuum and
delayed, radiatively driven emission-line variations. Combining $R_{{\rm
BLR}}$ with the emission-line velocity full-width-half-maximum $V_{\rm
FWHM}$,
assumed to be gravitationally determined and derived from the variable
line profile, virial masses $M$ follow from $M=fV_{\rm FWHM}^{2}R_{{\rm
BLR}}/G$, with $f$ a factor of order unity that depends on the detailed
emission-line gas distribution (and which may be determined by measuring
the velocity-dependent emission-line response). The results of Wandel et
al.\ (1999), Kaspi et al.\ (2000), and Peterson et al.\ (2000) are based
on these reverberation techniques. The magnitude and wavelength-dependence
of any continuum time delays, when combined with the spectral energy
distribution, may be used to constrain the product $M\dot{M}$ under
certain conditions (Collier et al.\ 1999), hence $\dot{M}$ follows given
$M$ from the associated emission-line reverberation measurement. The width
of the continuum delay distribution at many wavelengths and the velocity
dependent emission-line response (e.g., Welsh \& Horne 1991) may be used
to measure the source inclination. These methods for measuring $\dot{M}$
and $i$ have not been applied with much success to existing datasets,
since they lack sufficient signal-to-noise and monitoring duration. Other
methods for measuring $i$ include accretion disk fitting to Fe K$\alpha$
(e.g., Nandra et al.\ 1997) and combined UV continuum and \Hbeta\
measurements (Rokaki \& Boisson 1999).

We undertook a program of coordinated multiwavelength observations of the
NLS1 galaxy Akn~564 ($z=0.0247$, Huchra, Vogeley, \& Geller 1999) in the
summer of 2000. Akn~564 is the brightest known NLS1 galaxy in the
2--10\,keV band, with hard X-ray luminosity $L_{{\rm X}} \simeq
10^{43.4}\,{\rm erg}\,{\rm s}^{-1}$ (Turner et al.\ 2001).  It has a steep
power-law continuum ($\Gamma _{\rm hard} \approx 2.5$), ionized Fe K line
features, and a steep soft excess, as described by Turner et al. (2001).  
A 35 day \ASCA\ observation shows evidence for flux variations of
$\approx$ 36\%
(Turner et al.\ 2001, Edelson et al.\ 2001). Optical monitoring by
Giannuzzo et al.\ (1998) reveals \Hbeta\ variations of about 8\% on time
scales of about 4 years, rather lower than observed in other S1 galaxies
over similar periods (e.g., about 25\% for NGC 5548). Optical spectra
reveal the presence of strong [\caii] and \feii\ emission (van Groningen
1993). No program of UV monitoring has been previously undertaken on this
source, but \S{3} presents an archival IUE observation. Our program
included \ASCA\ (Turner et al.\ 2001 hereafter, Paper I), \HST\ (this
paper), optical (Shemmer et al.\ 2001 hereafter, Paper III), and \RXTE\
(Pounds et al.\ 2001) observations, and represents the most comprehensive
contemporaneous multiwavelength study of a NLS1 galaxy to date. In this
paper, we present the UV continuum and emission-line variability results
obtained with {\em HST} during the period 2000 May~9 to July~8.  These
observations were primarily intended to measure the effective size of the
UV BLR of a NLS1, and permit measurement of a virial mass $M$ based on
multiple emission lines, and thereby further constrain the nature of NLS1
galaxies. Akn~564 also has a rich UV absorption spectrum (Crenshaw et al.\
1999), and discussion of our absorption-line results will be deferred to a
later paper. In \S{2}, we describe the observations. We discuss the
properties of the mean spectra in \S{3}. The continuum and emission-line
variability are presented in \S{4} and \S{5}, respectively. Our results
are discussed and summarized in \S{6} and \S{7}, respectively.

\section{Observations}

We observed the nucleus of Akn~564 with the Space Telescope Imaging
Spectrograph (STIS) on the {\em Hubble Space Telescope} ({\em HST}) on 46
occasions during the period 2000 May~9 to July~8. The first five visits
were separated by intervals of $\sim$ 5 days. Beginning on the fifth
visit (2000 May~29), the sampling interval was decreased to $\sim$ 1 day
for the remaining monitoring period.  We obtained low-resolution spectra
during each visit with the G140L and G230L gratings, thereby providing
full UV coverage at a spectral resolution of approximately 1.2\,\AA\ over
the range 1150--1730\,\AA\ and 3.2\,\AA\ over the range
1570--3150\,\AA. We used the $52''\times 0\arcsecpoint5$ slit for the
low-resolution spectra to maximize throughput and ensure accurate absolute
photometry. As Akn~564 is a point source in the STIS spectral images, there was no significant resolution degradation by using the wide slit.
The exposure times were 1200 and 520 seconds for each G140L and G230L
spectrum, respectively, except on 2000 May~29, when the respective
exposure times were 1434 and 816 seconds. To estimate the effects of the
intrinsic absorption on the UV emission-line profiles, we obtained
high-resolution echelle spectra of the nucleus on 2000 May~29 with the
E140M grating at a spectral resolving power of $\lambda/\Delta\lambda
\approx 45,000$. The E140M spectra were obtained through the
$0\arcsecpoint2 \times 0\arcsecpoint2$ aperture during four consecutive
orbits to yield a combined exposure time of 10,310 sec. The spectra were
reduced using the IDL software developed at NASA's Goddard Space Flight
Center for the STIS Instrument Definition Team (Lindler 1998). For the
low-dispersion spectra, we used standard point-source processing with an
extraction height (perpendicular to the dispersion) of $0\arcsecpoint275$
to obtain flux-calibrated G140L and G230L spectra as a function of
wavelength.

Small wavelength calibration uncertainties exist between spectra of a
given grating. To remove these effects, relative wavelength
calibration of spectra for each grating is required. The methodology
we employ is analogous to that described by Korista et al.\ (1995)
(except we do not subtract a continuum fit from each spectrum). We
cross-correlate the spectra near the emission-line peaks of
\Lya\,$\lambda1216$, \civ\,$\lambda1549$, and \heii\,$\lambda1640$ 
for the G140L grating, and near the peaks of
\ciii]\,$\lambda1909$ and \mgii\,$\lambda2798$ for the
G230L grating. Hence, we determine the nearest whole
pixel shifts for each spectrum relative to its appropriate mean
spectrum. The majority of the spectra (about 90\%) required either a
one or zero-pixel shift, and the remaining spectra required a two-pixel
shift. The uncertainty in our relative wavelength calibration is of
order $\pm0.5$ pixels, i.e., about 0.6\,\AA\ and 1.7\,\AA\ for the G140L
and G230L gratings, respectively. We made no attempt to intercalibrate
the G140L and G230L spectra, given their difference in resolution.

\section{Mean Spectra}

The observed mean and root mean square (rms) of the 46 G140L and G230L
spectra are presented in Figures 1 and 2, respectively. The rms
spectra have been corrected for the bias due to measurement errors.
Furthermore, these
spectra have been corrected for
Galactic reddening using $E(B-V)=0.06$ mag. (Schlegel et al.\ 1998). We
note narrow line Balmer decrement and
\heii\,$\lambda1640$/\heii\,$\lambda4686$ ratios suggest internal
reddening is potentially large for Akn~564 (Walter \& Fink 1993; Paper
III), i.e., $E(B\!-\!V) \approx 0.2$\,mag. Determination of the
internal reddening in
Akn~564 will be discussed in a forthcoming paper, Crenshaw et al.\ 2001. 
The
mean
spectra are rich in both emission and absorption features. The
prominent emission features have been identified and labelled in
Figs.\ 1 and 2. Many of the unlabelled, poorer contrast features may
be associated with \feii\ emission. Intrinsic UV absorption in
\Lya\,$\lambda1216$, \nv\,$\lambda1240$, 
\siIV\,$\lambda1397$, and \civ\,$\lambda1549$ is present, along with
Galactic
absorption in lines such as  \siIV\,$\lambda1397$ and 
\mgii\,$\lambda2798$ (Crenshaw et al.\ 1999). 
We have made no attempt to identify the complex of
absorption features shortward of \Lya\,$\lambda1216$,
since this spectral region is affected by time-dependent residuals due
to imperfect subtraction of geocoronal \Lya\ emission.
We defer discussion of the rms spectra to \S{5}.

In order to quantify the prominent emission-line characteristics, we
determine pure emission-line spectra from which appropriate mean spectral
measurements are then made. This was done through cubic spline
interpolations over the absorption features and subtraction of a power-law
fit to the continuum. Our spectral measurements (to be discussed in this
section) assume zero internal reddening. The continuum is defined by four
nominally line-free bands chosen by visual inspection of the mean
spectrum, Fig. 1; 1155--1180\AA, 1350--1380\AA, 1460--1500\AA, and
1620--1660\AA. These bands only approximate the true continuum level,
since Balmer recombination continuum, \feii\ blends and other weak broad
emission contaminate them to some extent. The rms spectrum, Fig.\ 1,
indicates these bands are not notably biased by line variability. The
best-fit continuum is described by $F_{\lambda}=k(\lambda / 1000\,{\rm
\AA})^{\alpha}$ with $\alpha = -0.88\pm0.01$ and $k=(1.56\pm0.01) \times
10^{-14}\,{\rm ergs}\,{\rm s}^{-1}\,{\rm cm}^{-2}\,{\rm \AA}^{-1}$. The
continuum fit uncertainties are purely statistical and should be treated
with caution, since systematic errors due to reddening, for example, are
non-negligible. This UV continuum is redder than $F_{\lambda} \propto
\lambda ^{-1.34}$, typically observed in quasars (O'Brien et al. 1988).
This may be due to, for example, disk irradiation (assuming the UV
emission arises in an accretion disk) and/or internal reddening. We choose
to extrapolate this continuum into the G230L spectral range, since here
the continuum level is poorly defined, as the aforementioned contaminants
strengthen. We note, in this regime, the extrapolated continuum is about
30\% below the local, pseudo-continuum defined largely by the Fe II
blends. The validity of the above continuum fit awaits detailed
consideration of these Fe II blends and corrections for internal
reddening.

Our emission-line spectral measurements are detailed in Table 1. For each
emission line in column (1) we give the FWHM (FW$_{0.5}$) in column (2)
and flux contained therein (F$_{0.5}$) in column (3). Similarly, columns
(4) and (5) give the full width at 20\% maximum intensity (FW$_{0.2}$),
and flux therein (F$_{0.2}$). Column (6) gives the line centroid $\lambda
^{{\rm cen}}_{0.2}$, defined by F$_{0.2}$. No attempt has been made to
deconvolve blended lines, hence column (1) refers to the primary emission
line only. In Table 1, columns (4) through (6)  are empty for
\ciii]\,$\lambda1909$ and \mgii\,$\lambda2798$ because F$_{0.2}$ is
ill-defined,
\ciii]\,$\lambda1909$ is blended with \siIII]\,$\lambda1892$ and the
estimated, extrapolated continuum near \mgii\,$\lambda2798$ is
significantly below the observed level. We note that
all values in Table 1 are in the observed frame. We chose to measure line
fluxes at 20\% maximum intensity to minimize the effects of line blending,
by measuring the primary emission line flux only.

The uncertainties in the full width and flux measurements (Table 1) are
dominated by systematic errors attributable to uncertainties in the
applied absorption (where applicable) and reddening corrections. We
assess the uncertainty
associated with our absorption corrections by using cubic spline and
linear interpolations over the \Lya\,$\lambda1216$ absorption feature.
Figure 3 presents an illustrative example of the ambiguity in the
\Lya\,$\lambda1216$ emission line profile due to the different
interpolation schemes. We find uncertainties of $\sim$ 15\% and 8\% for
our line width and flux measurements, respectively. We do not consider the
systematic uncertainty in our applied reddening corrections, given the
aforementioned possibility of large internal reddening. For wavelengths
\gax 1800\AA, systematic errors of $\sim 30\%$ due to
uncertainties in our continuum fit may be important, but are not
considered further.

The \Lya\,$\lambda1216$, \nv\,$\lambda1240$, and \civ\,$\lambda1549$
emission lines have FWHMs of $\approx 2000$\,\kms, compared to typical
values of $\sim5000$\,\kms\ for S1 galaxies. Their FW$_{0.2}$ values are
$\approx 4000$\,\kms. For \heii\,$\lambda1640$, the FW$_{0.5}$ and
FW$_{0.2}$ estimates are $\approx$ 1000 and 3000 km\,s$^{-1}$,
respectively (from the mean G140L spectrum). 

The emission-line profiles appear symmetric about their systemic
wavelengths, but this is difficult to quantify due to contaminating
emission, such as \oiii\,$\lambda1664$ in the red wing of
\heii\,$\lambda1640$, and absorption features affecting many of the line
profiles. The centroids, $\lambda^{{\rm cen}}_{0.2}$, of the observed
emission-line fluxes of \Lya\,$\lambda1216$, \nv\,$\lambda1240$,
\siIV+\oiv]\,$\lambda\,1400$, \civ\,$\lambda1549$, \niii]\,$\lambda1750$,
and \mgii\,$\lambda2798$ are consistent with those expected based on the
systemic redshift $z=0.0247$, given the 1.2\AA\ and 3.2\AA\ spectral
resolutions of the G140L and G230L gratings. The \oi\,$\lambda1336$ and
\niv\,$\lambda1523$ emission-line flux centroids are redshifted and
blueshifted by 2.5\AA\ and 1.6\AA, respectively. The \oi\,$\lambda1336$
$\lambda^{{\rm cen}}_{0.2}$ is possibly biased by \sii\,$\lambda1340$
emission. The \heii\,$\lambda1640$
emission-line flux centroid is blueshifted by 1.3\AA\ ($\approx 230\,{\rm
km}\,{\rm s}^{-1}$). The \heii\,$\lambda1640$ $\lambda^{{\rm cen}}_{0.2}$
measured from the G230L mean spectrum is most likely biased by
\oiii\,$\lambda1664$ contamination. Figure 4 presents the mean
\Lya\,$\lambda1216$, \civ\,$\lambda1549$, and \mgii\,$\lambda2798$
emission line profiles as a function of velocity, with the above defined
continuum fit subtracted. The emission line
profile amplitudes have been normalized to unity. The high ionization
lines of \Lya\,$\lambda1216$ and \civ\,$\lambda1549$ and the low
ionization line of \mgii\,$\lambda2798$ do not appear notably blueshifted
or redshifted relative to systemic velocities. These results are discussed
in \S{6.1}.

We find some significant differences between emission-line flux ratios in
Akn~564 relative to those of more typical Seyfert 1 galaxies.  From Table
1, measured line ratios for Akn~564 are ${\rm (\civ/\Lya)} =
0.27\,(1.09)$,
${\rm (\ciii]/\Lya)}^* = 0.16\,(0.13)$, ${\rm (\ciii]/\civ)}^*=
0.65\,(0.12)$, ${\rm (\siIV + \oiv]/\civ)} = 0.48\,(0.07)$, ${\rm (\siIV +
\oiv]/\Lya)} = 0.13\,(0.08)$, ${\rm (\mgii/\Lya)}^* = 0.33\,(0.22)$, and
${\rm (\heii/\civ)} = 0.78\,(0.09)$. The line ratios in parenthesis
are
those for the S1 galaxy NGC~5548, taken from Clavel et al.\ (1991). Line
ratios with an asterisk denote those (for Akn~564) estimated using the
fluxes at 50\% maximum intensity, compared to 20\% for the others. This was
necessary because, as mentioned previously, F$_{0.2}$ is ill-defined for C
III]\,$\lambda\,{\rm 1909}$ and \mgii\,$\lambda2798$. The
uncertainties in these line ratios are dominated by systematic effects,
discussed above, and are of order 25\%. This error estimate includes an
$\sim$ 15\% bias due to excluding flux in the line-wings. For line ratios
that include \ciii]\,$\lambda1909$ and \mgii\,$\lambda2798$, the
error estimate
may be notably underestimated on account of the continuum fit
uncertainties
described above. Our emission-line ratios show
Akn~564 exhibits weaker \civ\,$\lambda1549$ and stronger
\siIV+\oiv]\,$\lambda\,1400$ emission compared to NGC~5548. These results
are consistent with earlier results of Wilkes et al.\ (1999) and
Kuraszkiewicz et al.\ (2000). We note our \mgii/\Lya\ ratio of 0.33 is
notably larger than the mean value of 0.05 derived from the small sample of
NLS1s by Kuraszkiewicz et al.\ (2000). This is probably
largely attributable to different
continuum level estimates near to \mgii\,$\lambda2798$.

Finally, we checked archival IUE observations of Akn~564 to assess whether
it was in a comparatively low or high flux state. Figure 5 presents the
IUE SWP observation of 1984 Jan 17. For comparison purposes, our mean
G140L
spectrum of Fig.\ 1 is overlaid as the thicker line. Both spectra have
been corrected for Galactic reddening as detailed above. The continuum and
emission-line fluxes are in qualitative agreement. In particular, the
\Lya\,$\lambda1216$ and \nv\,$\lambda1240$ line profiles are very similar;
and suggest negligible emission-line flux differences between the 1984 and
2000 observations. Other IUE observations from 1981 are also in
qualitative agreement with our observations.

\section{Continuum Variability Characteristics}


\subsection{Light Curves}

We select five nominal continuum bands by visual inspection of the mean
and rms spectra presented in Figs.\ 1 and 2: 1350--1380\AA, 1460--1500\AA,
1620--1660\AA, 2070--2130\AA, and 2960--3040\AA, henceforth referred to by
their mean observed wavelengths 1365\AA, 1480\AA, 1640\AA, 2100\AA, and
3000\AA, respectively. We re-iterate that each of these continuum bands
overestimates the true continuum level on account of various contaminants
(\S{3}). These bands are not notably biased by emission-line variability,
and thereby provide good approximations for the continuum variability at
these wavelengths. Light curves, describing the continuum variability
during the 60-day monitoring period, are presented in Figure 6, with the
continuum bands as labelled. We note our observed continuum flux
measurements are not corrected for reddening.

All the continuum regions show the same qualitative behavior. The 1365\AA\
variations may be described by; an increase in flux of $\sim$ 20\% during
the first 15 days, followed by a decrease of $\sim$ 16\% over the next 5
days. These variations are followed by two flares at about ${\rm
JD}-2450000 \approx 1697$ and 1711. The rising trough-to-peak variations
are about 18\% in $\sim$ 3 days, with similar declining peak-to-trough
variations occuring over longer time scales of several days. Both flare
events appear to be asymmetric, the later one more so and of longer
duration. Between these two events was a relatively quiescent period of
about 5 days. Thereafter, the variations are less pronounced, at about the
5\% level. For the longer wavelength variations, at 2100\AA\ and 3000\AA,
the rms variations are of reduced amplitude with broader asymmetric flare
profiles. The peaks of the 3000\AA\ flares are delayed, relative to those
at 1365\AA, by 1-2 days. The contemporaneous ASCA observations exhibit
larger amplitude correlated variations (including similar but narrower
twin flare events), possibly delayed by about 0.4 days (Papers I and III).
We
note the continuum variations may be slightly undersampled during the
intensive monitoring period of ${\rm JD}-2450000 \approx 1694$--1734, and
are definitely undersampled for prior times, given that $\sim $ 10\% flux
amplitude changes on time scales of 1 day are evident (see below).

We characterize the 1365--3000\,\AA\ variations by measuring two 
common variability parameters: the ratio of the maximum to minimum 
flux $R_{{\rm max}}$ and
the amplitude of the intrinsic variability relative to the mean flux
$F_{{\rm var}}$. The latter is corrected for the measurement errors 
$\varepsilon$, 
\begin{equation}
F_{{\rm var}}=\frac{1}{\overline{F}} \sqrt{(\sigma ^{2} _{{\rm F}} - 
\Delta ^{2} )} ,
\end{equation}
where $\overline{F}$, $\sigma _{{\rm F}}$, and $\Delta ^{2}=(1/N)
\sum_{i=1}^{N} \varepsilon ^{2}_{i}$ are respectively the standard mean
flux, rms flux, and mean square of the measurement errors, with $N=46$
the number of data points in the light curve (Edelson, Krolik, \& Pike
1990, Rodr\'{\i}guez-Pascual et al.\ 1997b). The results are presented in Table 2 for each
of the
light curves, as listed in column (1). Columns (2)--(5) give,
respectively, the mean flux $\overline{F}$, the rms flux $\sigma
_{{\rm F}}$, $F_{{\rm var}}$, and $R_{{\rm max}}$. Both $F_{{\rm
var}}$ and $R_{{\rm max}}$ are potentially biased by constant flux
components, e.g., \feii\ emission. However,
these effects are likely to be small. On time
scales of about 60 days, the fractional amplitude of the intrinsic
variations is
essentially constant between 1365\,\AA\ and 2100\,\AA\ at about 6\%,
with the full range of variations $R_{{\rm max}}-1 \approx 0.31$. At
3000\AA\ the fractional variability amplitude is about 4\%, with the full
range of variations $R_{{\rm max}}-1 \approx 0.19$. This
level of intrinsic UV variability is about a factor 3 less than
that found in typical S1 galaxies, which typically display fractional flux
variations $F_{{\rm var}} \approx$ 18\% on time scales of 60 days
(Collier, Peterson \& Horne 2001). 

The 1365--1640\,\AA\ continuum variations show evidence for $\sim$ 10\%
flux amplitude changes on time scales of about 1 day, as seen in the
events beginning at about ${\rm JD}-2450000 \approx 1696$, 1710, and 1707,
and suggest a fraction of the UV continuum emitting region must come from
a compact region of \lax 1 light day in size. The 2100--3000\,\AA\
variations show evidence for reduced flux amplitude changes of $\sim$ 5\%
on similar time scales. For comparison, in S1 galaxies far UV variations
of about 5\% rms occur on similar time scales (e.g., Korista et al.\
1995; Welsh et al.\ 1998). These faster, larger amplitude variations
suggest NLS1s (at least Akn~564) exhibit more variability power on short
(day) time scales. We confirm this through an autocorrelation analysis.
Figure 7 presents the Akn~564 1365\AA\ (solid line) and NGC~7469 (a S1
galaxy) 1315\AA\ (dotted line) autocorrelation functions (ACFs). The
relative steepness of the 1365\AA\ ACF (compared to the 1315\AA\ ACF)
indicates Akn~564's fluctuation power density spectrum is flatter than
that for NGC~7469, and thereby exhibits more power on short time scales.
We note Pounds et al.\ (2001) have shown that the X-ray variations in
Akn~564 are faster than those of typical S1 galaxies. The
full-width-half-maximum of the ACFs are 3.27 and 4.93 days for Akn~564 and
NGC~7469, respectively, and are indicative of characteristic UV
variability time scales. By assuming the mass ratio of the sources is
determined by the variability time scale ratio of 0.66, we estimate the
mass of Akn~564 to be $M \sim 5 \times 10^{6}\,\Msun$, given a mass
estimate for NGC~7469 of $8 \times 10^{6}\,\Msun$ (Wandel et al.\ 1999).
This mass estimate (for Akn~564) is in good agreement with that derived
from our emission-line reverberation results, discussed in \S{6.2}.

\subsection{Cross-Correlation Analysis}

The continuum light curves of Fig.\ 6 clearly display correlated
variations between 1365--3000\,\AA. The peaks and troughs of the flare
events at ${\rm JD}-2450000 \approx 1697$ and 1711 occur approximately
simultaneously; as mentioned above, the 3000\AA\ flare peaks appear
delayed by about 1-2 days relative to those at shorter wavelengths.  
Differences between the light curves are also evident, for example, decay
time scales for the two flares appear comparatively longer at wavelengths
$\geq$
2100\AA, and short (day) time scale 1365\AA\ variations appear washed out
at 3000\AA. In order to quantify the nature of the
correlations between
continuum variations we undertook a cross-correlation analysis. Two
algorithms were employed to compute
cross-correlation functions (CCFs), the interpolated CCF (ICCF) of
Gaskell \& Sparke (1986) as implemented
by White \& Peterson (1994), and the $Z$-transformed discrete
correlation function (ZDCF) algorithm of Alexander (1997). For our
data, the results recovered by the two algorithms are in agreement.


Figure 8 shows the CCFs obtained by cross-correlating each of the four
longer-wavelength continuum light curves with the 1365\,\AA\ light curve.  
The solid line and data points with error bars show the ICCF and ZDCF
CCFs, respectively, and are in good agreement. The 1365--3000\,\AA\
variations are highly correlated as evidenced by maximum values of the
cross-correlation coefficients $r_{\rm max} \approx 0.9$.  The
1480--2100\AA\ CCFs all peak at about zero lag, thereby suggesting these
continuum variations occur quasi-simultaneously. The 3000\AA\ CCF peaks
away from zero at about 0.5 days. The 2100\AA\ and 3000\AA\ CCFs appear
asymmetric, compared to the 1480\AA\ and 1640\AA\ CCFs, and suggest their
responses extend over a larger range of positive delays. Figure 9 presents
similar CCFs for four optical continuum regions; 4875--4915\AA,
5197--5237\AA, 6551--6622\AA, and 6935--7004\AA, henceforth referred to by
the observed wavelengths 4900\AA, 5200\AA, 6600\AA, and 6900\AA,
respectively (for consistency with Paper III). The optical data are from
Paper III. These CCFs have been computed as described above. The optical
variations are correlated with those at UV wavelengths with $r_{\rm max}
\approx 0.5$. The decreasing maximum correlation coefficient with optical
wavelength is likely due to progressively less coherent time delay smeared
responses and/or increasing noise dilution (see Table 4 in Paper III). The
optical CCFs clearly peak away from zero at about 2 days.

Table 3 summarizes our cross-correlation results.  Column (1) lists the
light curve that has been cross-correlated with the 1365\,\AA\ continuum
light curve. Columns (2) and (3) give the centroid of the CCF $\tau _{{\rm
cen}}$, as determined by the ICCF and ZDCF algorithms. The ICCF centroid
is calculated over all points above 0.8 times the maximum
cross-correlation coefficient $r_{{\rm max}}$, whereas the ZDCF centroid
is computed from all points near the peak with a cross-correlation
amplitude above half that of the peak. Columns 4 and 5 report the time
delay $\tau _{{\rm peak}}$, measured from the peak of the CCF. Columns 6
and 7 give $r_{{\rm max}}$, for the ICCF and ZDCF, respectively. Column 8
details the FWHM of the ICCF. The reported errors on $\tau _{{\rm peak}}$
and $\tau _{{\rm cen}}$ for the ICCF are the 1-$\sigma$ uncertainties as
determined by the model-independent quasi-bootstrap and flux randomization
method of Peterson et al. (1998). We find no evidence for lags between the
1365\,\AA\ and 1640\,\AA\ light curves, as evidenced by peak and centroid
lag measurements consistent with zero. The 2100\,\AA\ and 3000\AA\
variations, however, lag those at 1365\,\AA\ by $\sim 0.3$ and 1.0 day,
respectively. Here and hereafter, we note the centroid lags since they are
a less biased estimator of the size of the responding emission
region. These lags are greater than zero at 94\% and 99\%, respectively.
The optical variations at $\sim 5000$\,\AA\ lag behind those at 1365\,\AA\
by about 2 days at no less than 99\% confidence. The optical variations at
wavelengths longer than 6000\,\AA\ appear to lag those at 1365\,\AA\ by a
similar amount, although statistically the suggested lags are not
significantly different from zero on account of their very low
amplitude ($\sim 1$\%) variability; which is comparable to the measurement
errors.

We investigate these wavelength-dependent continuum lags further by
binning the 46 G140L and G230L spectra (covering 1150--3140\AA) into
40\,\AA\ bins, and forming light curves based on the total flux in each
bin. We cross-correlate (using the ICCF) each of these light curves with
the 1365\,\AA\ light curve. The results are presented in Figure 10. In the
top panel, the histogram plot shows the centroid lag for each bin as a
function of the bin wavelength, thus producing a ``lag spectrum''. The
corresponding values of $r_{\rm max}$ are shown in the lower panel.  
Again, the error bars for each bin in the upper panel were determined by
the method of Peterson et al.\ (1998). The solid line represents the
best-fit function $\tau \propto (\lambda ^{\gamma} -
\lambda_{0}^{\gamma})$, with $\lambda _{0}=1365\,{\rm \AA}$, $\gamma =
2.4\pm0.1$, and reduced chi-squared $\chi ^{2}_{\nu} = 0.07$ for $\nu =
50$ degrees of freedom, to the lag measurements. The small value of $\chi
^{2}_{\nu} << 1$ suggests the lag uncertainties may be overestimated.  
There is a clear trend of increasing lag with wavelength throughout the
UV: the variations at 3100\,\AA\ lag those at 1365\,\AA\ by about 0.9
days. The lag measurements for $\lambda \gtsim 2050$\,\AA\ are non-zero at
no less than 90\% confidence. For wavelengths shorter than 2050\,\AA, the
lag measurements are consistent with zero delay. The 1150--3140\,\AA\
variations are well correlated with $r_{{\rm max}} \approx 0.9$. In the
region where the G140L and G230L spectra overlap, i.e., $\sim$
1600--1700\AA, the lag measurements are in good agreement. The lag
spectrum is reasonably smooth with no clear positive deviations near the
emission-line wavelengths, as would be the case if the emission-line
response time scales were measurably larger. Similarly, there are no
negative dips in the maximum correlation coefficient spectrum, with the
possible exception near \mgii\,$\lambda2798$. These results support the
negligible presence of emission-line variability to be discussed in \S{5}.
We note the dip in the maximum correlation coefficient at about 1700\,\AA\
is due to calibration uncertainties in the G230L spectra. Similarly, the
slight depression in $r$ at \lax 1200\AA\ is due to calibration errors in
the G140L spectra. We note the two bins with the largest centroid lag
errors are those contaminated by \Lya\,$\lambda1216$ and
\mgii\,$\lambda2798$.

We use the optical lag measurements (Table 3) to extend the UV lag
spectrum. The UV/optical lag spectrum is presented in Figure 11. The solid
line represents the best-fit function $\tau \propto (\lambda ^{\gamma} -
\lambda_{0}^{\gamma})$, with $\lambda _{0}=1365\,{\rm \AA}$, $\gamma =
1.3\pm0.1$, and $\chi ^{2}_{\nu} = 0.09$ for $\nu = 54$ degrees of
freedom. The dotted line represents the best-fit function to the UV data
alone (as in Fig.\ 10), i.e., with $\gamma = 2.4\pm0.1$. By including the
optical data, the lag--wavelength relationship appears to flatten.

\section{Emission-Line Variability}

The largely featureless rms spectra of Figs.\ 1 and 2 immediately show
that any emission-line variability is of low amplitude. Figure 1 shows
\Lya\,$\lambda1216$, \nv\,$\lambda1240$, \siIV+\oiv]\,$\lambda\,1400$,
\civ\,$\lambda1549$, and \heii\,$\lambda1640$ variations are present with
amplitudes of $<$ 4, 5, 6, 5, and 4\%, respectively. Figure 2 indicates
\siIII\,$\lambda1892$, \ciii]\,$\lambda1909$, and \mgii\,$\lambda2798$
variations with amplitudes of $<$ 6, 6, and 4\%, respectively. For cases
where emission-line variations are present or suggested, the bulk of the
variation occurs in the core of the line. The variations at these
wavelengths, for emission lines contaminated by intrinsic absorption, may
be a superposition of both emission and absorption-line variability. A
discussion of the latter is deferred to a later paper. We note the
\nv\,$\lambda1240$ variations appears to be slightly asymmetric, with a
deficit of response in its blue wing. The variations at about 1215\AA\ are
spurious, and attributable to time-dependent residuals from imperfect
subtraction
of geocoronal \Lya\ emission.

Emission-line light curves were constructed by direct integration, summing
the flux in specified, absorption-free wavelength regions after
subtracting a continuum defined by a linear fit to the nearest continuum
regions on either side of the emission-line region. Our attempts at
validating and possibly improving our direct integration light curves
through spectral fitting with the IRAF task SPECFIT (Kriss 1994) failed.
This was because the spectral complexity of the emission-line profiles
permitted too much freedom in the fitting process such that artifical
emission-line variations were introduced by corresponding changes in the
adopted absorption profile. The blanket exclusion of any absorption
regions from our variable emission-line flux regions, defined by their
respective rms spectra (Figs.\ 1 and 2), severely inhibited our ability to
extract the already low-amplitude variations. We were only able to extract
significant emission-line variations for \Lya\,$\lambda1216$. The
compartively
smaller fluxes and larger measurement errors for the other emission-lines
precluding extraction of significant variations.

Figure 12 presents the \Lya\,$\lambda1216$ light curves for the monitoring
period. The data points with error bars describe the emission-line
variations between 1240--1243\,\AA\ and 1247--1250\,\AA\ (``light curve
1''), and the dashed line those between 1240--1250\,\AA\ (``light curve
2''); the latter includes any possible variable contribution from the
intrinsic \hi\ absorption feature.  In both cases the continuum is defined
by a linear fit between 1155--1180\,\AA\ and 1350--1380\,\AA. The 1365\AA\
light curve, \S{4.1}, is scaled and vertically shifted to fit light curve
1, and is shown by the solid line. We note our observed emission-line flux
measurements are not corrected for reddening.

The \Lya\,$\lambda1216$ variations resemble those of the continuum, but
with greatly reduced amplitude and appear delayed by about a few days. In
each case, the
flux increases by $\sim 3$\% during the first 15 days, and similar flare
events occur at similar times, i.e., ${\rm JD}-2450000 \approx 1697$ and
1711. The contrast of any features in the light curves is poor on account
of the low-amplitude variability, which is comparable to
the
measurement uncertainties. On time scales of about 60 days, the intrinsic
line variations are $\sim 1$\%, with the full range of variation $\sim
7$\%. Moreover, the \Lya\,$\lambda1216$ light curves are clearly well correlated and
exhibit similar variability patterns. These results suggest any intrinsic
\hi\
absorption-line variability does not notably affect the emission-line
variations. 

Cross-correlation of the \Lya\,$\lambda1216$ and 1365\AA\ variations
quantifies their apparent similarity. Our results are presented in Figure 
13 and Table 3. Figure 13 presents \Lya\,$\lambda1216$ CCFs for both line
light curves detailed above. The solid line and filled data points
represent the ICCF and ZDCF CCFs, respectively, as described earlier, for
light curve 1, and the dashed line and open data points are for light
curve 2. The \Lya\,$\lambda1216$ variations are correlated with the
1365\,\AA\ variations with a maximum cross-correlation amplitude $r_{{\rm
max}} \approx 0.5$. The probability of exceeding $r \approx 0.5$ in a
random sample of observations (of the same number, $N=45$, as here) drawn
from an uncorrelated parent population is about 0.1\% (Bevington \&
Robinson 1992), thereby the \Lya\,$\lambda1216$ variations are correlated
with those at 1365\AA\ at about 99.9\% confidence. The formal significance
of this correlation is most likely over-estimated, since not all
correlated data points are independent. Detailed simulations will be
required in order to quantify the magnitude of this bias. By directly
cross-correlating light curves 1 and 2, we confirm they are correlated
with essentially zero time delay. The peak and centroid lag measurements
for light curves 1 and 2 are consistent to within the large uncertainties.
Moreover, for light curve 1 the lag measurements are consistent with zero
delay. Our results only permit an upper limit for the size of the
\Lya\,$\lambda1216$ emission region, $R_{{\rm BLR}}^{{\rm Ly\alpha}}$ \lax
3 light days. We caution, the low amplitude \Lya\,$\lambda1216$
variations may indicate the bulk of the emission region is at larger  
radii, as discussed in \S{6.2}.

\section{Discussion}

\subsection{Nature of the UV Broad Line Region}

A complete study of the nature of the UV BLR of Akn~564 requires detailed
photoionization modelling of the spectral energy distribution and is
beyond the scope of this paper. Here, we offer a preliminary discussion
of our results in the context of previous work. 

The BLR of Akn~564 is characterised by smaller gas velocity dispersions
(as determined by the emission-line velocity full-width-half-maximum) than
found in
typical S1 galaxies (\S{3}), as noted earlier by Rodr\'{\i}guez-Pascual et
al.\ (1997a), i.e., about 2000\,\kms compared to 5000\,\kms. These authors
found evidence for similar magnitude UV line widths in about a dozen
NLS1s; see also
Kuraszkiewicz et al. (2000). The full widths of the low and
high-ionization lines are comparable, see Fig.\ 4, and interestingly
similar to those of the \niii]\,$\lambda1750$ and \ciii]\,$\lambda1909$
intercombination lines. For example, the full widths of
\heii\,$\lambda1640$, \niii]\,$\lambda1750$, \ciii]\,$\lambda1909$, and
\mgii\,$\lambda2798$ are 1800, 1900, 1900, and 1700 km\,s$^{-1}$,
respectively (Table 1; as measured from the G230L spectra). 

As noted in \S{3}, the broad emission-line profiles of Akn~564 are
approximately symmetric
about their systemic wavelengths. These
results, and those above, contrast with UV/optical
observations of quasars (e.g., Gaskell 1982; Wilkes 1984; Corbin 1991) and
S1 galaxies, showing that high-ionization lines are shifted blueward of
low-ionization lines, which are approximately at systemic redshifts, and
slightly broader. Leighly (2000) presented observations of two NLS1s (IRAS
13224-3809 and 1H 0707-495) which also showed the high ionization lines
were strongly blueshifted and much broader than the low ionization
\mgii\,$\lambda2798$ line, similar to results for the NLS1 galaxies 1 Zw 1
and NGC~4051 presented by Laor et al.\ (1997) and Peterson et al.\ (2000),
respectively. Collinge et al.\ (2001) find the \civ\,$\lambda1549$ line in
NGC~4051 to be blueshifted by about 100 km\,s$^{-1}$. These prior results
are often interpreted as evidence for BLR radial motions. The low
ionization lines may arise in an accretion disk, with the high ionization
lines emitted in a wind launched from this disk (e.g., Collin-Souffrin et
al.\ 1988). Our results suggest an absence of outflowing high ionization
BLR gas with velocities \gax 230 km\,s$^{-1}$ in Akn~564. Moreover, the
approximate symmetry of the emission-line profiles may indicate the
predominance of optically thin broad line emission, since it is emitted
essentially isotropically and traces the full velocity field (Netzer 1976;
Davidson 1977; Shields, Ferland, \& Peterson 1995). 

The level of intrinsic line variability is notably less than observed in
typical Seyfert 1 galaxies over comparable time scales (\S{5}). For
example, NGC~5548 (Korista et al.\ 1995) showed \Lya\,$\lambda1216$,
\civ\,$\lambda1549$, and \heii\,$\lambda1640$ variations of $\sim13$\%,
$\sim9$\%, and $\sim14$\%, respectively. Similarly, the same lines in
NGC~7469 (Wanders et al.\ 1997) showed variations of $\sim10$\%,
$\sim9$\%, $\sim14$\%. This difference may simply be a consequence of the
reduced level of assumed ionizing continuum variations combined with
geometrical dilution due to the finite BLR size. To investigate this
scenario we convolved the observed continuum variations with thin shell
broad line regions at radii of 6 and 2 light days (i.e., the predicted
sizes of the \Lya\,$\lambda1216$ and \heii\,$\lambda1640$ emission
regions, see \S{6.2}), respectively. In these cases, we find
\Lya\,$\lambda1216$ and \heii\,$\lambda1640$ variations of order 3 and
6\%, respectively. These results suggest this type of scenario does not
completely explain the reduced level of emission-line variability in
Akn~564. On the other hand, the line-emitting gas in Akn~564 may be fully
ionized in hydrogen, and hence optically thin to the Lyman continuum. This
would lead to greatly reduced emission-line variability, with detectable
variations occurring only for sufficiently large continuum variations that
induce significant changes in the ionization structure of the emission
region. Similarly, this scenario alone is likely insufficient to account
for the lack
of line variability across the observed range of ionization, as evidenced
by the calculations of Ferland \& Mushotzky (1982). However, we note these
calculations are more appropriate for S1 BLRs, and thereby may be
misrepresentative of NLS1s. Some combination of both the above
effects appears a reasonable explanation for the reduced level of
emission-line variability in Akn~564.

The unusual UV spectral properties of Akn~564 (\S{3}), combined with
stronger ionizing fluxes relative to other S1 galaxies, suggest NLS1s BLRs
have a physically distinct nature from those found in S1 galaxies and
quasars. Kuraszkiewicz et al. (2000) posit NLS1 BLRs characterized by gas
cloud densities $n_{{\rm H}} \approx 10^{11}\,{\rm cm}^{-3}$, ionization
parameters $U \approx 0.01$, and BLR radii $R_{{\rm BLR}}^{{\rm UV}}
\approx 20$ light days. On the other hand, Rodr\'{\i}guez-Pascual et al.
(1997a)
posit $n_{{\rm H}} \approx 10^{7.5}\,{\rm cm}^{-3}$, $U \approx 0.3$, and
BLR radii $R_{{\rm BLR}}^{{\rm UV}} \approx 2$ light days; under these
conditions the broad line emitting gas becomes fully ionized in hydrogen,
and thereby optically thin to the Lyman continuum. Our emission line
profiles and variability results hint at a predominance of optically thin
broad line emitting gas in Akn~564. In combination with a small estimated
BLR size, \S{5}, of \lax 3 light days, these conditions appear more
readily
reconcilable with those suggested by Rodr\'{\i}guez-Pascual et al.\
(1997a), as noted above.

\subsection{Evidence for a Black-Hole Accretion Disk System}

The wavelength-dependent continuum time delays of \S{4.2} suggest a
stratified continuum reprocessing region extending a couple of light days
from the central black hole, and provide possible evidence for an
accretion-disk structure (Collier et al.\ 1998). Simple irradiated
accretion disk models with radial temperature profiles $T \propto
R^{-3/4}$ predict $\tau \propto\lambda ^{4/3}$. Flatter temperature
profiles predict steeper $\tau$--$\lambda$ relationships, and are
realizable with, e.g., irradiated flared accretion-disk models. These
models may be able to explain the steeper UV lag spectrum presented in
\S{4.2}. The fine details are determined by the geometry and structure of
the accretion disk, relative prominence of viscous heat dissipation and
irradiation effects, and the emission physics. Vaughan et al.\ (1999)
present evidence, based on X-ray data, for a strongly ionised disk in
Akn~564. Given the close proximity of the UV and X-ray emitting regions
(\S{4} and Papers I and III), our results support their conclusions. In
the case of NGC~7469, the only other source with a clear detection of
wavelength-dependent continuum lags ( Wanders et al.\ 1997, Collier et
al.\ 1998, Kriss et al.\ 2000), the UV/optical lag spectrum was consistent
with $\tau \propto \lambda^{4/3}$, and is similar to that observed in
\S{4.2}.  On the other hand, these continuum lags may be a result of
diffuse continuum emission from broad-line emitting clouds (Korista \&
Goad 2000), or some combination of both these effects.

In \S{5} we estimated an upper limit of 3 light days for the size of the
\Lya\,$\lambda1216$ broad line emitting region, $R_{{\rm BLR}}^{{\rm
Ly\alpha}}$. We may compare this estimate with that expected on the basis
of results obtained for other AGNs by using the Balmer emission region
size $R_{{\rm BLR}}^{{\rm H\beta}}$--luminosity relationships of Kaspi et
al.\ (2000) and Peterson et al.\ (2000). The luminosity $\lambda
L_{\lambda}(5100\,{\rm \AA})=\lambda F_{\lambda}(5100\,{\rm \AA}) 4 \pi
D^{2} \approx 2.4 \times 10^{43}\,{\rm erg}\,{\rm s}^{-1}$, adopting a
mean flux $F_{\lambda}(5100\,{\rm \AA}) = 4 \times 10^{-15}\,{\rm
erg}\,{\rm s}^{-1}\,{\rm cm}^{-2}\,{\rm \AA}^{-1}$ and a distance
$D=cz/H_{0} = 99$\,Mpc (assuming $H_{0} = 75\,{\rm km}\,{\rm s}^{-1}\,{\rm
Mpc}^{-1}$). This leads to an estimate for the size of the \Hbeta-emitting
region $R_{{\rm BLR}}^{{\rm H\beta}} \approx 12$ light days from the
$R_{{\rm BLR}}^{{\rm H\beta}}$--luminosity relationship shown in Fig.\ 6
of Peterson et al.\ (2000). From previous monitoring programs on S1
galaxies (Netzer \& Peterson 1997), we can estimate the size of the
\Lya\,$\lambda1216$ emitting region to be $R_{{\rm BLR}}^{{\rm Ly\alpha}}
\approx 0.5 R_{{\rm BLR}}^{{\rm H\beta}}$.  Similarly, we expect that the
highest ionization lines, \nv\,$\lambda1240$ and \heii\,$\lambda1640$,
arise in a region of extent $\sim 0.2 R_{{\rm BLR}}^{{\rm H\beta}}$. Based
on the optical luminosity of Akn~564, we thus predict $R_{{\rm BLR}}^{{\rm
Ly\alpha}} \approx 6$\,light days and $R_{{\rm BLR}}^{{\rm \nv}}\approx 2$
light days. The factor of $\sim$ 2 difference between our measured and
predicted values for $R_{{\rm BLR}}^{{\rm Ly\alpha}}$ is probably not
significant, since the intrinsic scatter in the $R_{{\rm BLR}}^{{\rm
H\beta}}$--luminosity relationship is of order a factor 5. These
results give us
confidence our $R_{{\rm BLR}}^{{\rm Ly\alpha}}$ estimate is in the right
ballpark.

By combining our $R_{{\rm BLR}}^{{\rm Ly\alpha}}$ upper limit of 3
light days
with
the emission-line velocity full-width-half-maximum $V_{\rm FWHM} \approx
2500\,{\rm km}\,{\rm s}^{-1}$,
derived from the rms emission-line profile (and assumed to be gravitationally
determined), we estimate a virial mass
upper limit from
\begin{equation} 
M=fV_{\rm FWHM}^{2}R_{{\rm BLR}}^{{\rm Ly\alpha}}/G\,^{<}_{\sim}\, 8 \times
10^{6}\,\Msun, 
\end{equation} 
where for consistency with Wandel et al.\ (1999) and Kaspi et al.\ (2000)
we use $f=3/ \sqrt{2}$. This mass estimate is consistent with the
independent
mass estimate of $M \sim 1 \times 10^{7}\,\Msun$ of Pounds
et al.\ (2001), based on a fluctuation power spectrum analysis of X-ray
variability in Akn~564. And is consistent with our mass estimate based on the
variability arguments of \S{4.1}. Furthermore, we note that
the position of
Akn~564 in the AGN mass--luminosity plane defined by Fig.\ 7 of Peterson et
al.\ (2000), is consistent with the best-fit regression line based on the
NLS1 galaxies alone, see Figure 14. This is consistent with the hypothesis
that
Akn~564
harbors a comparatively small black hole accreting with a higher accretion   
rate and/or is viewed more face-on than typical S1 galaxies; as found earlier
by Pounds et al.\ (2001).

We conclude with a cautionary note concerning our derived upper limits for
the size of Akn~564's \Lya\,$\lambda1216$ BLR and
black hole mass. The reduced continuum to
\Lya\,$\lambda1216$ variations ratio of $\sim$ 4, relative to $\sim$
1 for the S1 galaxies NGC~5548 and NGC~7469 (references as given in
\S{6.1}), suggests our \Lya\,$\lambda1216$ reverberations arise from a
region only about 1/4 of the size of the \Lya\,$\lambda1216$ region
probed by reverberation in S1 galaxies; and is therefore biased low. To
illustrate the potential importance of this bias, assume that the
\Lya\,$\lambda1216$ BLR is a thin shell of radius $R$. Equal
time delay intervals define equal areas on the shell, and if our
\Lya\,$\lambda1216$ response from $<$ 3 light days is from only 1/4 of
the
area of the shell, we underestimate $R$ by a factor of 4. This would lead 
to an $R_{{\rm BLR}}^{{\rm Ly\alpha}}$ upper limit of 12 light days. The
bias in
our black hole mass upper limit depends on the unknown nature of the gas
velocity field, and is not readily quantifiable without better data.

\section{Summary}

An intensive two-month monitoring program on the NLS1 galaxy Akn~564 was
undertaken with {\em HST} during 2000 May 9 to July 8. We summarize
our results as follows.

\begin{enumerate}

\item The fractional variability amplitude of the UV continuum variations
between
1365--3000\,\AA\ is about 6\% on time scales of 60 days. This level of
variability is about a factor of three less than that found in S1
galaxies. We find evidence for fast, large amplitude continuum variations,
e.g., trough-to-peak flux changes of $\sim18$\% in about 3 days, atypical
of those found in S1 galaxies that display similar amplitude variations on
longer time scales.

\item We present evidence for wavelength-dependent continuum time delays.
The continuum variations at 3000\,\AA\ lag behind those at 1365\,\AA\ by
about 1 day. By combining the UV data with the optical data described
in Paper III, we find that the variations at 5200\,\AA\ lag behind those
at 1365\,\AA\ by about 2 days.  These delays may be interpreted as
evidence for a stratified continuum reprocessing region, possibly an
accretion-disk structure. However, the delays may be a result of diffuse
continuum emission from broad-line emitting clouds, or some combination of
both these effects.

\item The \Lya\,$\lambda1216$ emission line exhibits intrinsic rms
variations of about 1\% on time scales of 60 days. These variations lag
those at 1365\AA\ by \lax 3 days, and combining this with the line width
yields a putative black hole mass limit of \lax $8 \times 10^{6}\,\Msun$.
This calculation assumes the line width is determined by the gravity of
the black hole. We caution, the low amplitude \Lya\,$\lambda1216$
variations may indicate the bulk of the emission region is at larger
radii, and thereby the veracity of our black hole mass limit is
questionable. The possible bias in our black hole mass estimate is
uncertain given the unknown nature of the gas velocity field. Despite the
unreliability of our mass estimate, it is consistent with the independent
estimate $M \sim 1 \times 10^{7}\,\Msun$ of Pounds et al. (2001), based
on a fluctuation power spectrum analysis of X-ray variability in Akn 564.

\item The root-mean-square spectra suggest other strong emission line
variability, e.g., of \civ\,$\lambda1549$ and \heii\,$\lambda1640$, occurs
with rms amplitudes of $<$ 5\%. The low level of NLS1 emission-line
variability is in contrast to that found in typical S1 galaxies, which
display rms flux variations of $\sim10$\% on similar time scales.

\end{enumerate}

\acknowledgements{We are pleased to acknowledge support for this work by
NASA through grant number HST--GO--08265.01--A from the Space Telescope
Science Institute, which is operated by the Association of Universities
for Research in Astronomy, Inc., under NASA contract NAS5--26555. This
research has made use of the NASA/IPAC Extragalactic Database (NED) which
is operated by the Jet Propulsion Laboratory, California Institute of
Technology, under contract with the National Aeronautics and Space
Administration. SM acknowledges NASA grant NAG5-8913 (LTSA). We are
grateful to an anonymous referee for useful comments.}

\clearpage

\begin{figure}

\caption{The observed mean (top panel) and root-mean-square (rms: bottom 
panel)
spectra of the 46 G140L spectra. These
spectra have been corrected for
Galactic reddening using $E(B-V)=0.06$ mag. The mean spectrum displays
many 
emission and
absorption-line features, and the solid line denotes the best-fit
power-law continuum, see text for further details. The continuum bands
used
in this paper are as indicated. The rms spectrum shows evidence
for \Lya\,$\lambda1216$, \nv\,$\lambda1240$, \siIV+\oiv]\,$\lambda\,1400$,
\civ\,$\lambda1549$, and \heii\,$\lambda1640$ variations with
amplitudes of $<$ 4, 5, 6, 5, and 4\%, respectively.}

\end{figure}
\
\begin{figure}

\caption{The observed mean (top panel) and root-mean-square (rms: bottom
panel) spectra of the
46 G230L spectra. These
spectra have been corrected for
Galactic reddening using $E(B-V)=0.06$ mag. The mean spectrum displays
many emission and 
absorption-line
features. The prominent emission-lines are marked, including the 
continuum bands
utilized in \S{4.1}. The rms spectrum shows evidence for
\siIII\,$\lambda1892$, \ciii]\,$\lambda1909$, and \mgii\,$\lambda2798$  
variations with amplitudes of $<$ 6, 6, and 4\%, respectively.}

\end{figure}

\begin{figure}
\caption{An illustrative example of the ambiguity in the
\Lya\,$\lambda1216$ emission-line profile, and hence spectral line
measurements (for which absorption corrections have been made) of \S{3},
due to different interpolation schemes. The solid line
histogram presents the \Lya\,$\lambda1216$ emission-line profile, and the
dashed
and dotted histograms detail the linear and cubic spline interpolation
corrections over
the absorption feature.}
\end{figure}

\begin{figure}
\caption{The mean \Lya\,$\lambda1216$,
\civ\,$\lambda1549$, and \mgii\,$\lambda2798$ emission-line profiles of
Figs.\ 1 and 2 as a
function of velocity. The emission line
profile amplitudes have been
normalized to unity. The high ionization lines of \Lya\,$\lambda1216$ and
\civ\,$\lambda1549$ and the low ionization line of \mgii\,$\lambda2798$ do
not appear notably blueshifted or redshifted relative to systemic
velocities.}
\end{figure}

\begin{figure}
\caption{The 1984 Jan 17 IUE SWP observation of Akn~564.
For
comparison purposes, our mean G140L spectrum of Fig.\ 1 is overlaid as the
thicker line. These
spectra have been corrected for
Galactic reddening using $E(B-V)=0.06$ mag. The continuum and
emission-line fluxes are in qualitative
agreement.}

\end{figure}

\begin{figure}
\caption{The observed continuum light curves for the 60 day monitoring
period, with mean observed wavelength as labelled. The amplitude of
the intrinsic rms variations is about 6\%, and they all exhibit the same
qualitative behaviour.}
\end{figure}

\begin{figure}
\caption{The Akn~564 1365\AA\ (solid line) and NGC~7469 ( a
Seyfert 1 galaxy) 1315\AA\ (dotted line)
autocorrelation functions (ACFs), respectively. The relative steepness of
the
1365\AA\ ACF (compared to the 1315\AA\ ACF) indicates Akn~564's fluctuation
power density spectrum is flatter than that for NGC~7469, and thereby  
exhibits more power on short time scales. The full-width-half-maximum of
the ACFs are 3.27 and
4.93 days for Akn~564 and NGC~7469, respectively, and are indicative of
characteristic UV variability timescales.}
\end{figure}

\begin{figure}

\caption{CCFs for four UV continuum regions, mean observed wavelengths as
labelled. The solid line and data points with error bars detail the ICCF
and ZDCF CCFs, respectively, and are in good agreement. The
1480--3000\,\AA\ variations are strongly correlated with those at 1365\AA\
as evidenced by maximum cross-correlation coefficients of $r_{{\rm max}}
\approx 0.9$. The 1480--2100\,\AA\ variations occur quasi-simultaneously,
since their CCFs all peak at about zero lag. The 3000\AA\ CCF peaks at
about 0.5 days.}
\end{figure}

\begin{figure} 
\caption{CCFs for four optical continuum regions, nominal observed
wavelengths as labelled. All CCFs have been computed with respect to the
variations at 1365\,\AA. The solid line and data points with error bars
detail the ICCF and ZDCF CCFs, respectively, and are in good agreement.
The 4900--6900\AA\ variations are correlated with those at 1365\AA, as
evidenced by maximum correlation coefficients of $r_{{\rm max}} \approx
0.5$. Moreover, they are delayed by about 2 days with respect to those at
UV wavelengths.}
\end{figure}

\begin{figure} 
\caption{The UV lag spectrum binned to 40\,\AA. Top panel: The solid
histogram denotes the centroid lags. The error bars are those derived 
from
a model-independent quasi-bootstrap and flux randomization method. The
solid line represents the best-fit function $\tau \propto (\lambda
^{\gamma}-\lambda_{0}^{\gamma})$, with $\lambda_{0}=1365\,{\rm \AA}$ and
$\gamma = 2.4\pm0.1$. There is a clear trend of increasing lag with
wavelength. Bottom panel: The solid histogram denotes the maximum
correlation coefficient. The 1150--3140\,\AA\ variations are well 
correlated
with $r_{{\rm max}} \approx 0.9$. The dip in the maximum
correlation coefficient at about 1700\,\AA\
is due to calibration uncertainties in the G230L spectra. See text for
further details.}
\end{figure}

\begin{figure} 

\caption{The UV/optical lag spectrum. The optical lag measurements,
denoted by the open symbols, are
derived from contemporaneous optical monitoring data presented in
Paper III. The solid line represent represents the best-fit function
$\tau \propto (\lambda ^{\gamma}-\lambda_{0}^{\gamma})$, with
$\lambda_{0}=1365\,{\rm \AA}$ and $\gamma = 1.3\pm0.1$. The dotted line
represents the best-fit function to the UV data, denoted by the
solid histogram, detailed in Fig.\ 10, i.e.,
$\gamma = 2.4\pm0.1$. By including the optical data, the lag-wavelength
relationship flattens.}
\end{figure}

\begin{figure}
\caption{\Lya\,$\lambda1216$ light curves for the 60 day monitoring
period.
The data points with error bars describe the emission-line variations
between 1240--1243\AA\ and 1247--1250\AA\ and the dashed line those
between 1240--1250\AA. In both cases the continuum is defined by a linear
fit between 1155-1180\AA\ and 1350--1380\AA. The 1365\AA\ light curve,
\S{4.1} and Fig.\ 6, is scaled and vertically shifted to fit light curve
1, and is shown by the solid line. The emission-line variations denoted 
by the dashed line, include any possible variable contribution from
the intrinsic \hi\ absorption feature. The intrinsic \Lya\,$\lambda1216$   
flux amplitude variations are about 1\%, and both light curves exhibit     
similar qualitiative behaviour which in turn mimics those of the continuum 
regions. We note the dashed line variations have been normalized to have   
the same mean value as those denoted by the data points.}
\end{figure}

\begin{figure}
\caption{CCFs for the \Lya\,$\lambda1216$ emission-line. The solid line
and filled data points
represent the ICCF and ZDCF CCFs, respectively for
light curve 1, and the dashed line and open data points are for light
curve 2 (see Fig.\ 12). The \Lya\,$\lambda1216$ variations are correlated
with the
1365\,\AA\ variations with a maximum cross-correlation amplitude $r_{{\rm
max}} \approx 0.5$. Their CCFs peak away from zero at about 2 days, and
the differences between them are not statistically significant.}
\end{figure}

\begin{figure}

\caption{The reverberation-based virial mass and optical luminosity
relationship for AGNs. The filled circles are Seyfert galaxies from Wandel
et al.\ (1999), and the open circles are QSOs from Kaspi et al.\ (2000).
The large triangles are the AGNs from the same sources whose broad lines
have widths less than $\sim$ 2000 km\,s$^{-1}$. Our upper limit mass
estimate for Akn~564 is shown as a filled diamond. The best-fit regression
line based on the narrow-line objects is denoted by the dashed line, with
$M \propto L^{0.48\pm0.08}$.}
\end{figure}

\clearpage

\begin{deluxetable}{lccccc}
\tablecolumns{6}
\tablecaption{Emission-Line Characteristics}
\tablewidth{0pt}
\tablehead{
\colhead{Line} & \colhead{FW$_{0.5}^{a}$} &
\colhead{F$_{0.5}^{b}$} & \colhead{FW$_{0.2}^{a}$} &
\colhead{F$_{0.2}^{b}$} & \colhead{$\lambda ^{{\rm cen}}_{0.2}$ (\AA)} \nl
\colhead{(1)} & \colhead{(2)} & \colhead{(3)} &  \colhead{(4)} &
\colhead{(5)} & \colhead{(6)} \nl
}
\startdata
\Lya\ &  2114  & 5.30 & 3553 & 6.51  & 1245.2 \nl
\nv\       &  2809  & 1.98 & 4725 & 2.56  & 1270.5 \nl
\oi\      &  788   & 0.15 & 1563 & 0.20  & 1338.1 \nl
\siIV+\oiv] &  3270  & 0.70 & 4859 & 0.84  & 1433.9 \nl
\niv\      &  928   & 0.35 & 1487 & 0.44  & 1521.6 \nl
\civ\      &  1934  & 1.33 & 3469 & 1.76  & 1586.3 \nl
\heii\      &  1195  & 0.91 & 2585 & 1.38  & 1679.6 \nl
\heii$^{c}$      &  1831  & 1.04 & 3875 & 1.44  & 1682.7 \nl
\niii]$^{c}$     &  1882  & 0.56 & 3930 & 0.80  & 1794.0 \nl
\ciii]$^{c}$     &  1920  & 0.86 & ------ & ------  & ------ \nl
\mgii$^{c}$      &  1659  & 1.73 & ------ & ------  & ------ \nl
\enddata
\tablenotetext{a}{In \kms.}
\tablenotetext{b}{In $10^{-13}$ ergs s$^{-1}$
cm$^{-2}$.}
\tablenotetext{c}{Spectral measurements from the G230L mean spectrum. 
All other
measurements from the G140L mean spectrum.}
\tablenotetext{d}{The spectra are corrected for a Galactic reddening of 
$E(B-V)=0.06$, and assume zero internal reddening.}
\end{deluxetable}
\clearpage

\begin{deluxetable}{lcccc}
\tablecolumns{6}
\tablecaption{Continuum and Emission-Line Variability Characteristics}
\tablewidth{0pt}
\tablehead{
\colhead{Light Curve} & \colhead{$\overline{F}\,^{\rm a} $} &
\colhead{$\sigma\,^{\rm a} _{{\rm F}}$} & \colhead{$F_{{\rm var}}$} &
\colhead{$R_{{\rm max}}$} \nl
\colhead{(1)} & \colhead{(2)} & \colhead{(3)} &  \colhead{(4)} &
\colhead{(5)} \nl
}
\startdata
1365\,\AA\ & 7.21 & 0.44 & 0.061 & 1.31 \nl
1480\,\AA\ & 6.99 & 0.46 & 0.065 & 1.33 \nl
1640\,\AA\ & 6.98 & 0.46 & 0.064 & 1.33 \nl
2100\,\AA\ & 5.82 & 0.30 & 0.051 & 1.28 \nl
3000\,\AA\ & 6.48 & 0.25 & 0.037 & 1.19 \nl
L$\alpha$\,$\lambda\,{\rm 1216}^{\rm b}$ & 16.95 & 0.30 & 0.015 & 1.07 \nl
L$\alpha$\,$\lambda\,{\rm 1216}^{\rm c}$ & 29.10 & 0.42 & 0.012 & 1.06 \nl
\enddata
\tablenotetext{a}{\raggedright Units are 10$^{-15}$ 
erg cm$^{-2}$ s$^{-1}$${\rm 
\AA}^{-1}$
for continuum fluxes and 10$^{-14}$ erg
cm$^{-2}$ s$^{-1}$ for the line fluxes. All light curves have $N=46$ 
data
points.}
\tablenotetext{b}{\raggedright \Lya\,$\lambda1216$ variations between
1240--1243\,\AA\ and 1247--1250\,\AA. See \S{5} for further details.}
\tablenotetext{c}{\raggedright \Lya\,$\lambda1216$ variations between
1240--1250\,\AA. See \S{5} for further details.}
\end{deluxetable}
\clearpage

\begin{deluxetable}{lcccccccc}
\tablecolumns{8}
\tablecaption {Cross-Correlation Results \label {cross-correlations}}
\tablewidth{0pt}
\tablehead{
\colhead{Band} & \multicolumn{2}{c}{$\tau_{{\rm cen}}^{a}$} &
\multicolumn{2}{c}{$\tau_{{\rm peak}}^{a}$} &
\multicolumn{2}{c}{$r_{{\rm max}}$} & \colhead{FWHM$^{a}$} \nl
\colhead{} & \colhead{ICCF} & \colhead{ZDCF} &  \colhead{ICCF} &
\colhead{ZDCF} & \colhead{ICCF} & \colhead{ZDCF} &\colhead{ICCF} \nl
\colhead{(1)} & \colhead{(2)} & \colhead{(3)} & \colhead{(4)} & 
\colhead{(5)} & \colhead{(6)} & \colhead{(7)} & \colhead{(8)} \nl
}
\startdata
1480\AA\ &  -0.0$^{+0.2}_{-0.2}$ & 0.0 & 0.0$\pm$0.0 & 
0.0$\pm$0.4
&
1.0 & 1.0
& 3.3
\nl
1640\AA\ &  0.1$^{+0.2}_{-0.2}$ & 0.2 & 0.0$\pm$0.1 &
0.0$\pm$0.4 & 0.9 & 1.0
& 3.0
\nl
2100\AA\ &  0.3$^{+0.4}_{-0.2}$ & 0.3  & 0.3$^{+0.2}_{-0.1}$ &
0.0$^{+0.4}_{-0.4}$ &
0.9 & 0.9
& 3.0
\nl
3000\AA\ &  1.0$^{+0.4}_{-0.3}$ & 0.7  & 0.5$^{+1.1}_{-0.0}$ &
0.0$^{+1.4}_{-0.4}$ &
0.8 & 0.7
& 3.5
\nl
4900\AA\ &  1.8$^{+0.5}_{-0.4}$ & 1.8  & 2.2$^{+0.2}_{-1.0}$ &
2.7$\pm$1.3 &
0.7 & 0.6
& 3.6
\nl
5200\AA\ &  1.8$^{+0.6}_{-0.4}$ & 1.7  & 1.4$^{+1.0}_{-0.1}$ &
1.7$^{+2.5}_{-1.0}$ &
0.6 & 0.5
& 3.7
\nl
6600\AA\ &  2.2$^{+1.7}_{-8.3}$ & 2.3  & 2.5$^{+1.8}_{-8.8}$ &
2.1$^{+0.8}_{-1.5}$ &
0.4 & 0.5
& 3.8
\nl
6900\AA\ &  2.6$^{+1.6}_{-5.1}$ & 2.4  & 3.0$^{+1.8}_{-6.4}$ &
2.7$^{+1.3}_{-1.8}$ &
0.4 & 0.3
& 3.7
\nl

L$\alpha$\,$\lambda\,{\rm 1216}^{b}$ & 0.9$^{+0.4}_{-1.2}$ & 0.4 &
0.7$^{+0.6}_{-1.1}$ &
1.0$^{+0.6}_{-1.7}$ &
0.6 & 0.5
& 5.0
\nl

L$\alpha$\,$\lambda\,{\rm 1216}^{c}$ & 2.8$^{+0.0}_{-2.3}$ & 1.9 &
2.7$^{+0.0}_{-2.2}$ &
1.0$^{+1.8}_{-1.4}$ &
0.5 & 0.4
& 3.6
\nl

\enddata
\tablenotetext{a}{In days.}
\tablenotetext{b}{\Lya\,$\lambda1216$ variations between
1240-1243\AA\ and 1247-1250\AA. See \S{5} for further details.}
\tablenotetext{c}{\Lya\,$\lambda1216$ variations between
1240-1250\AA. See \S{5} for further details.}
\end{deluxetable}
\clearpage

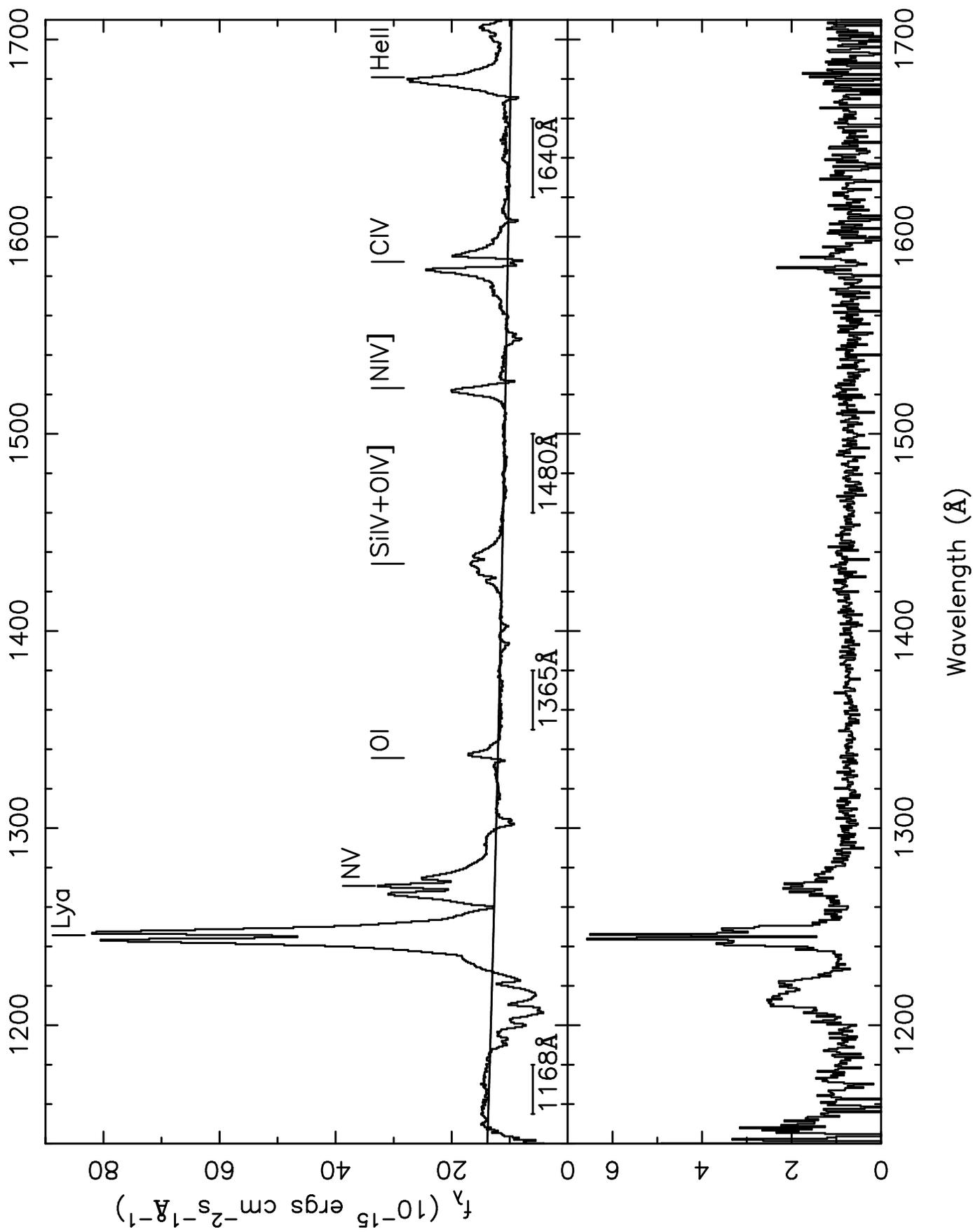
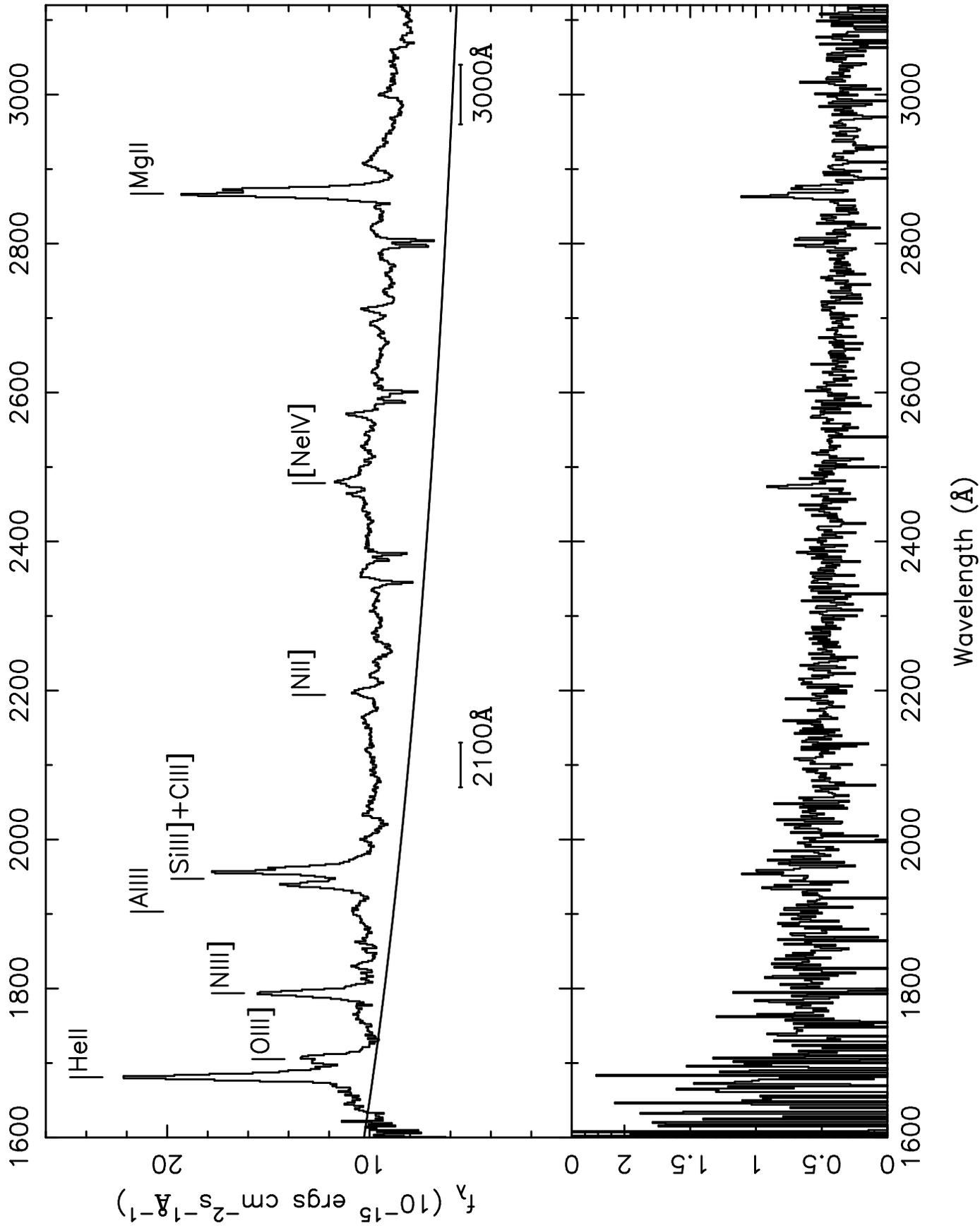
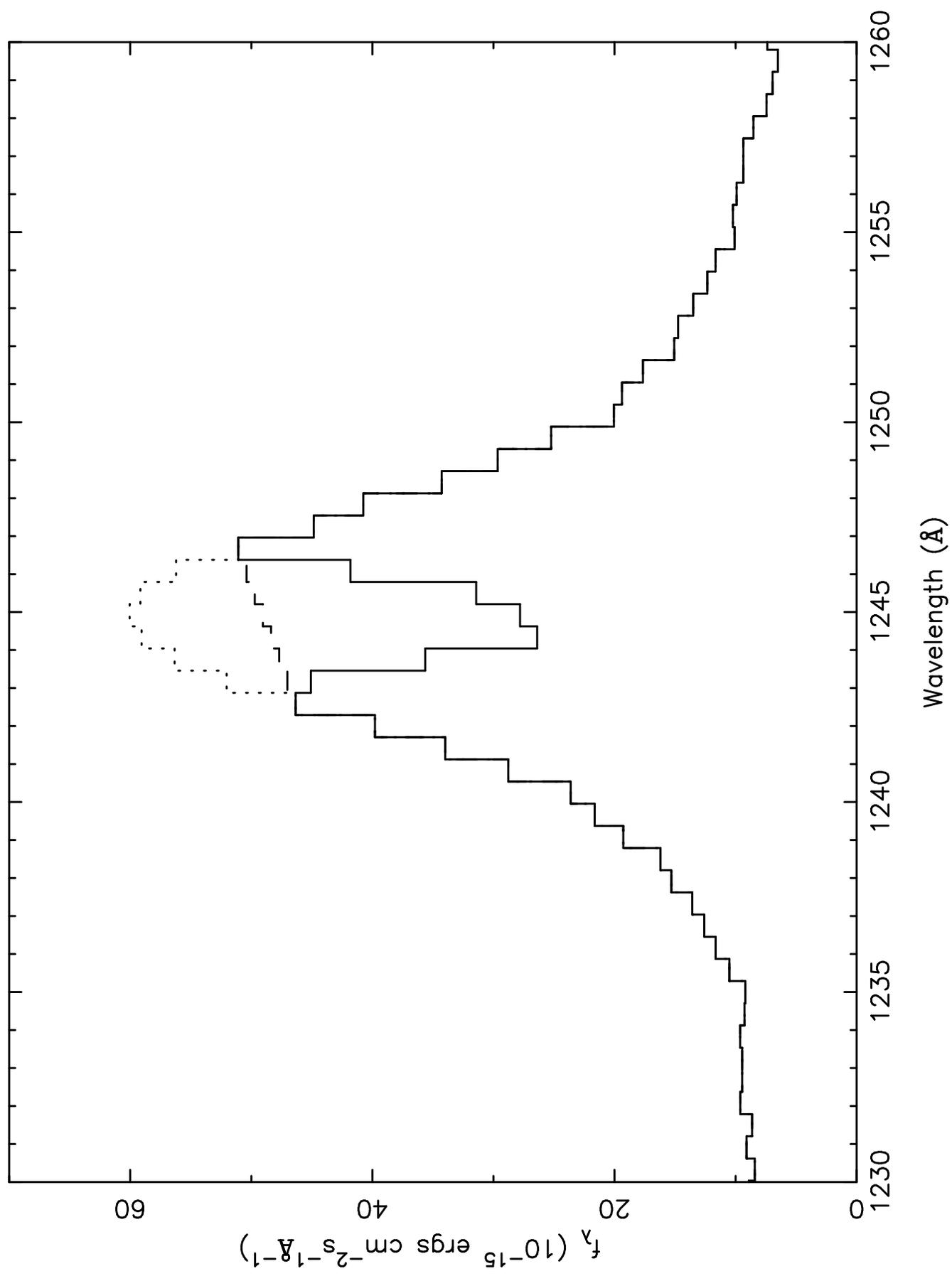
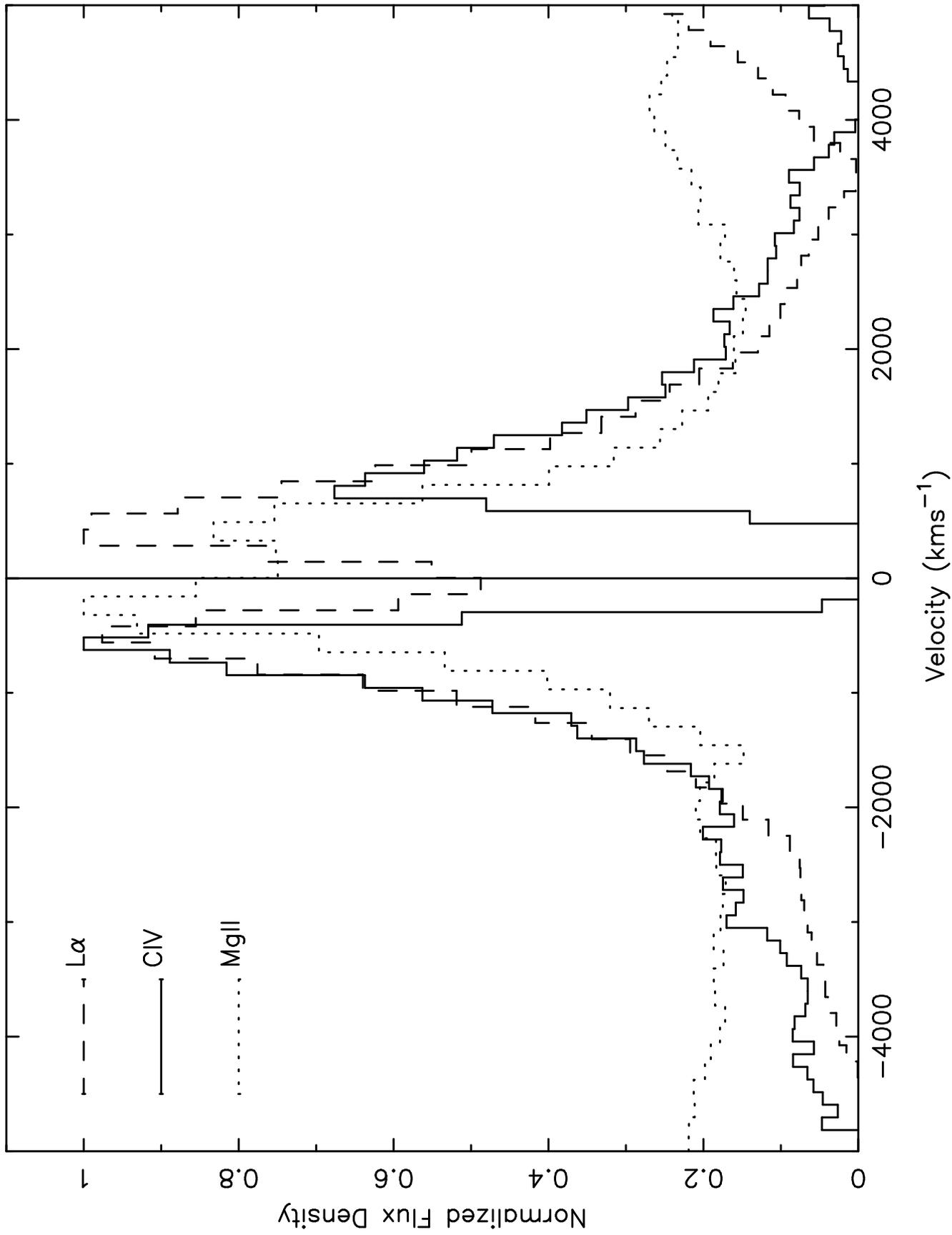
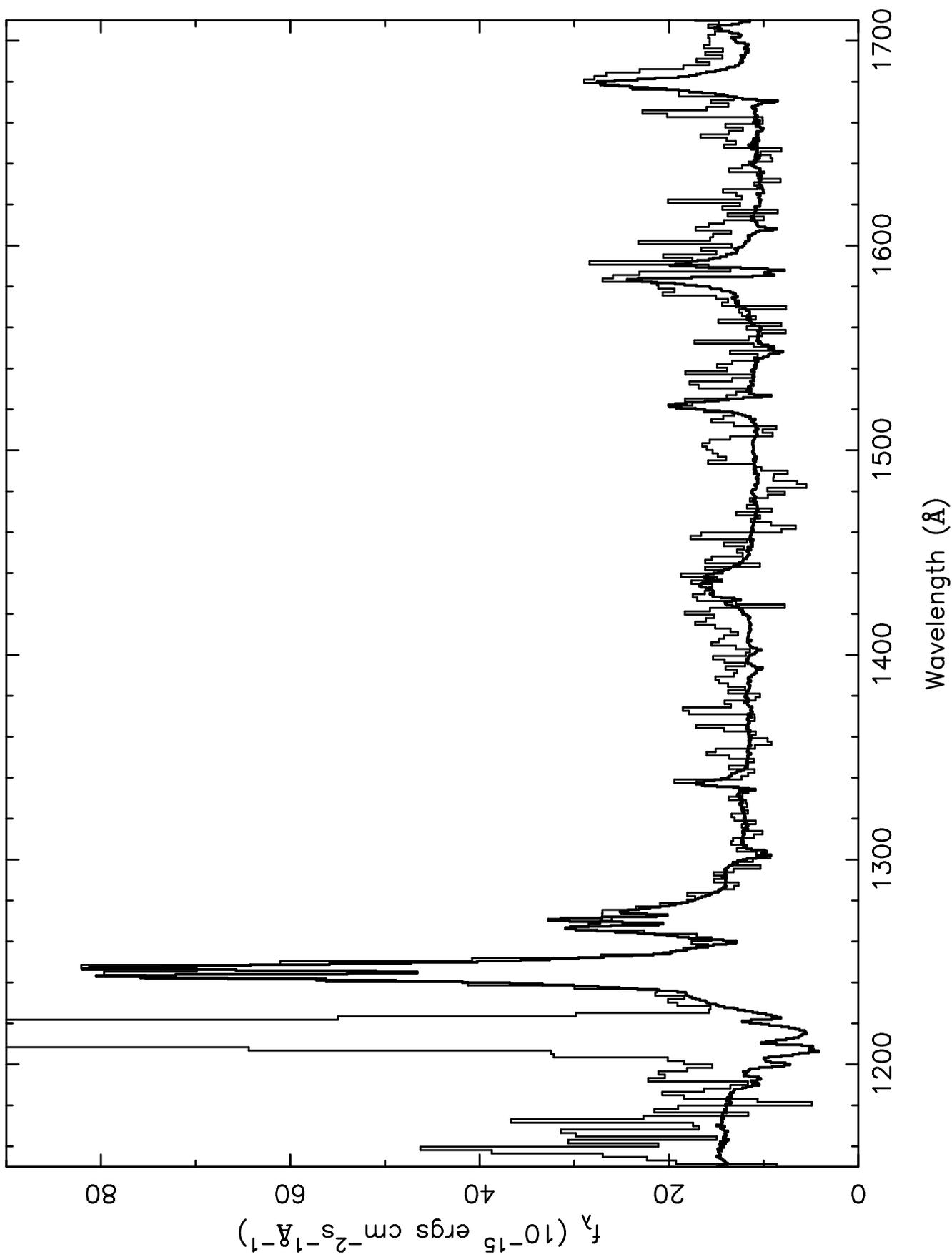
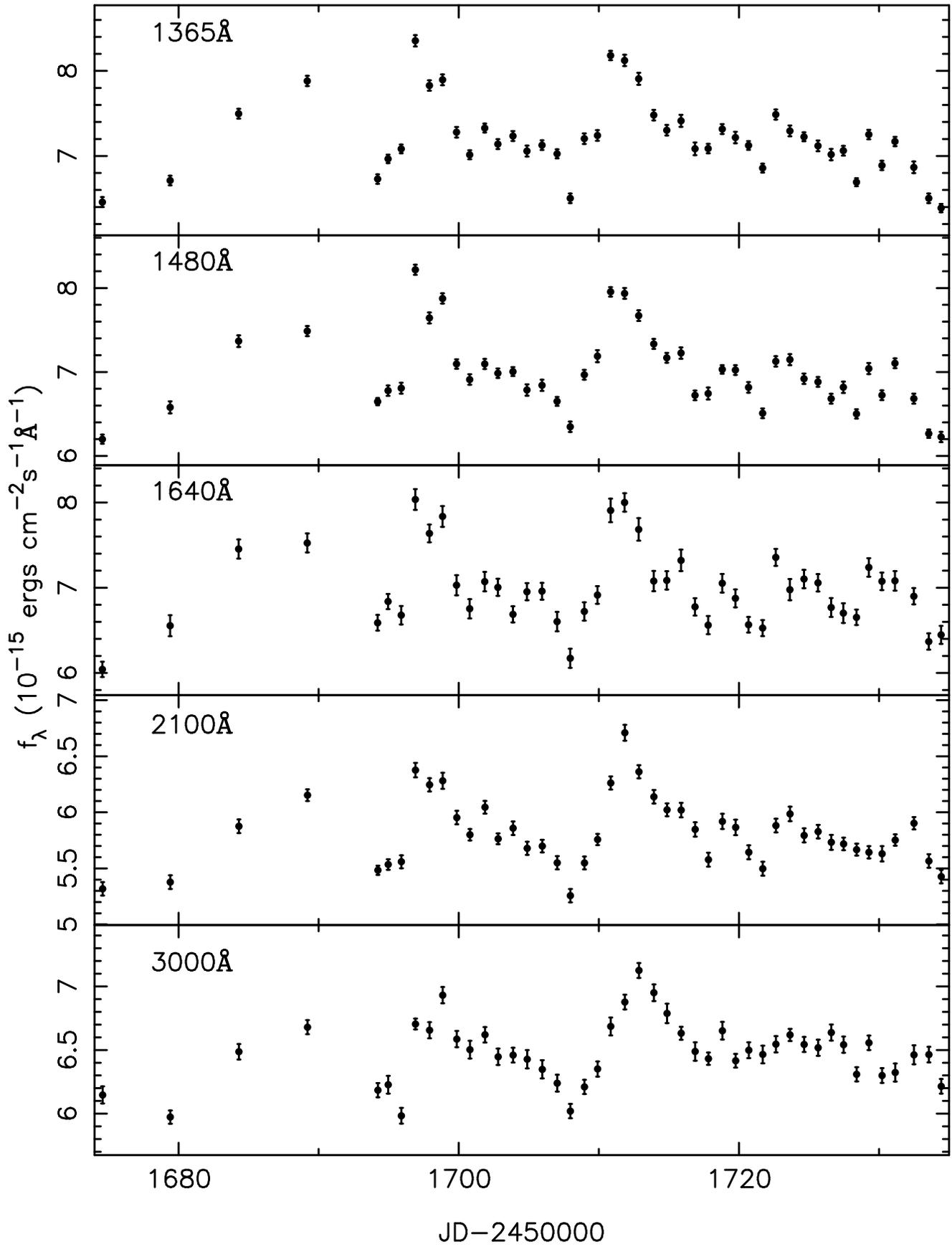
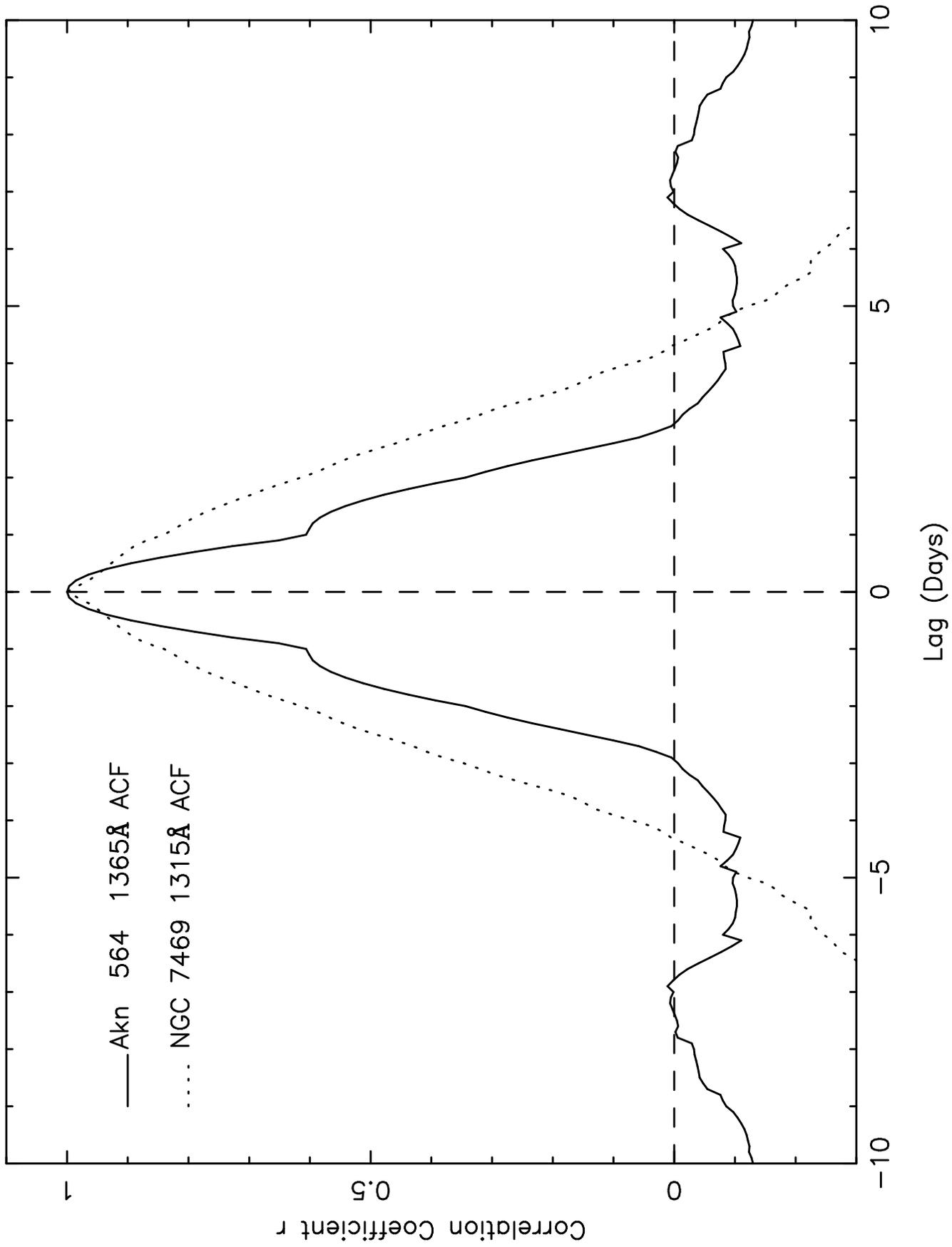
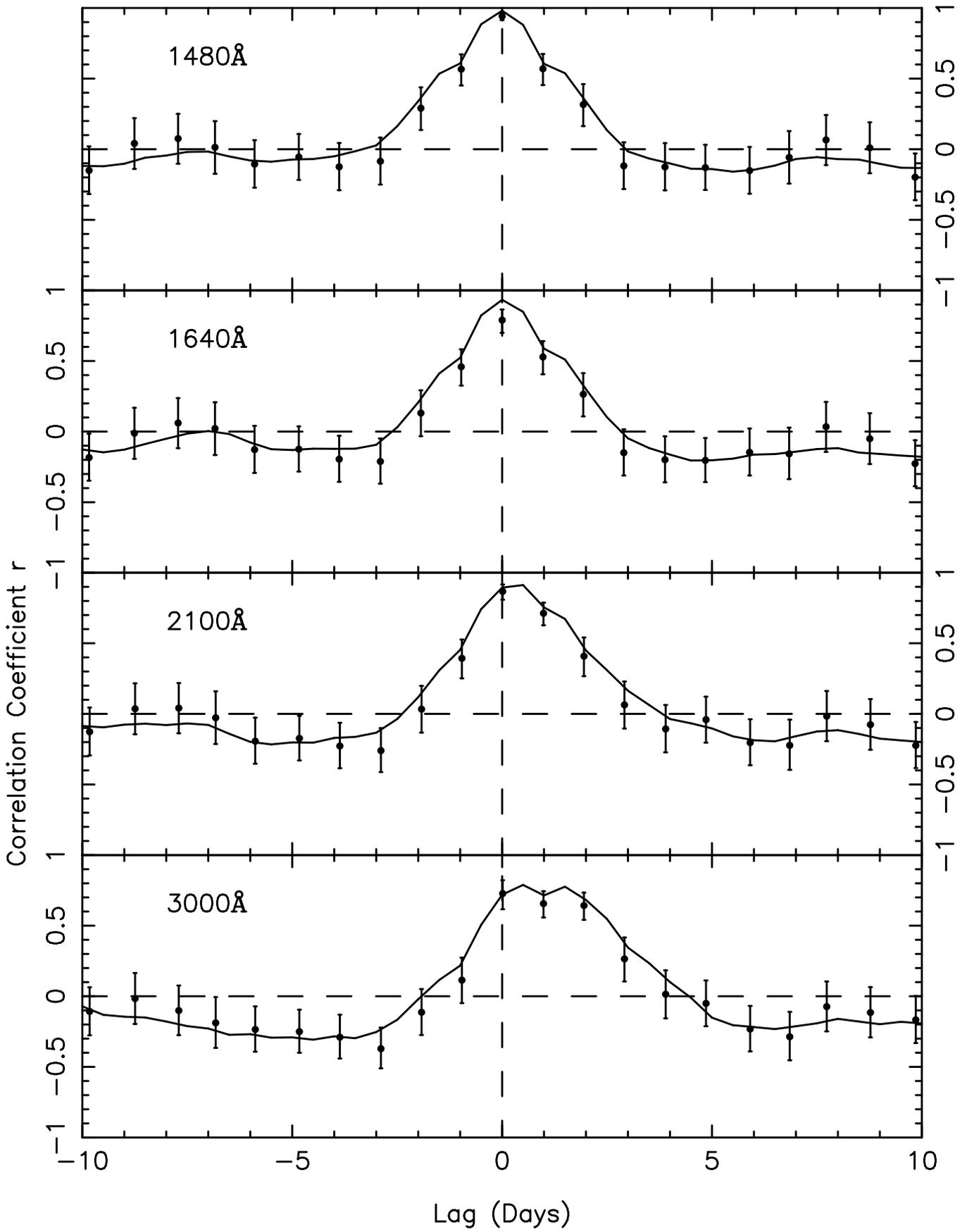
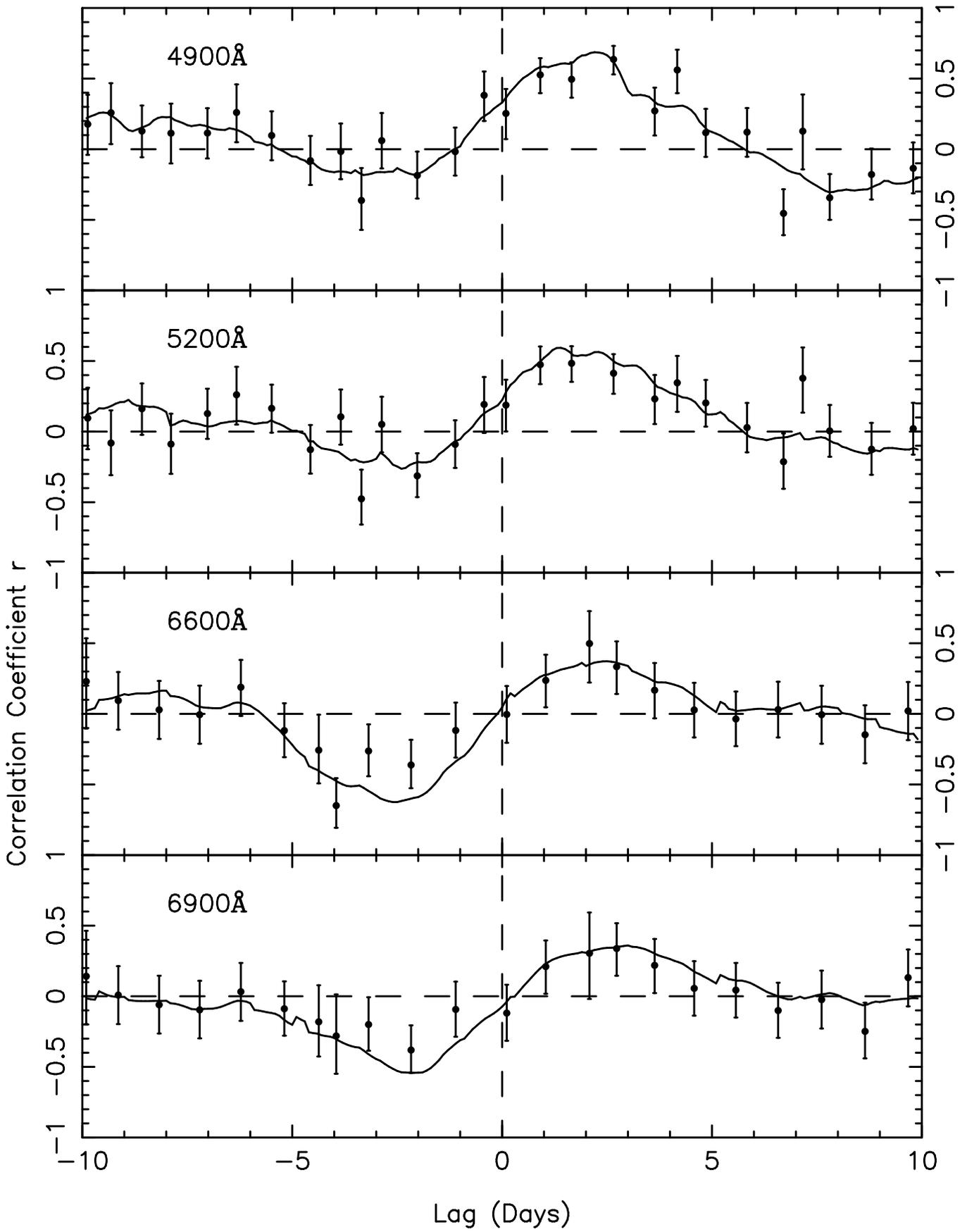
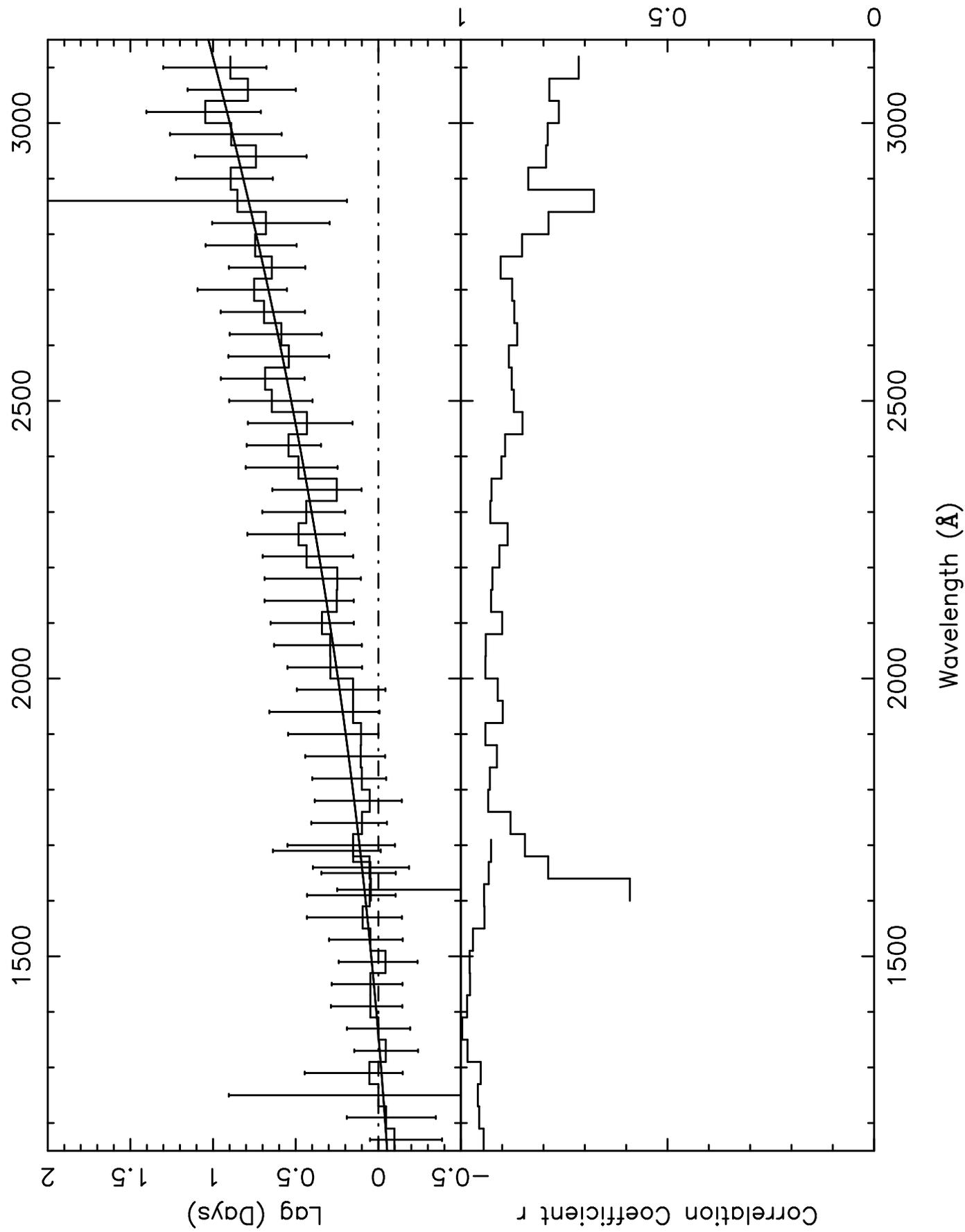
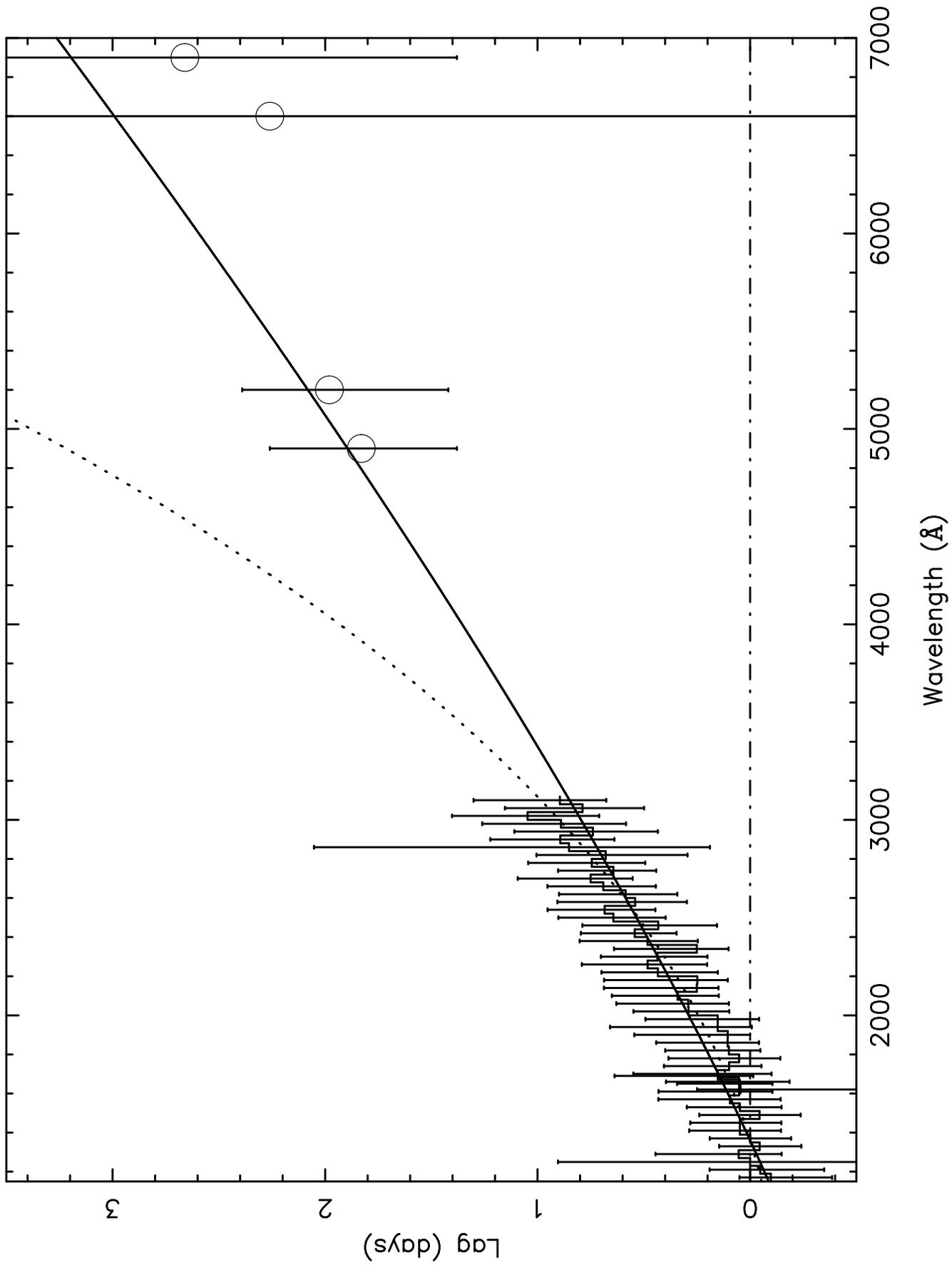
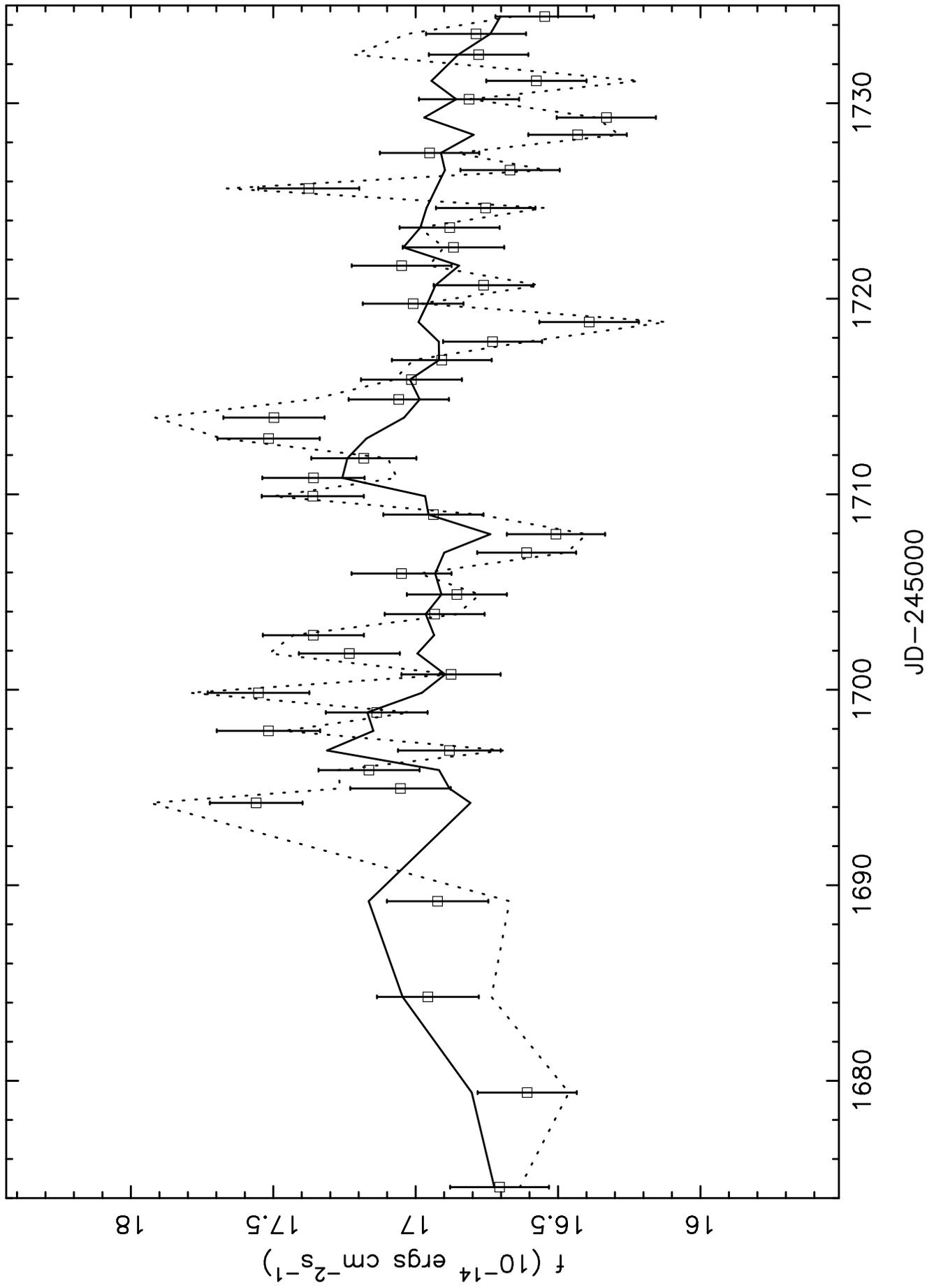
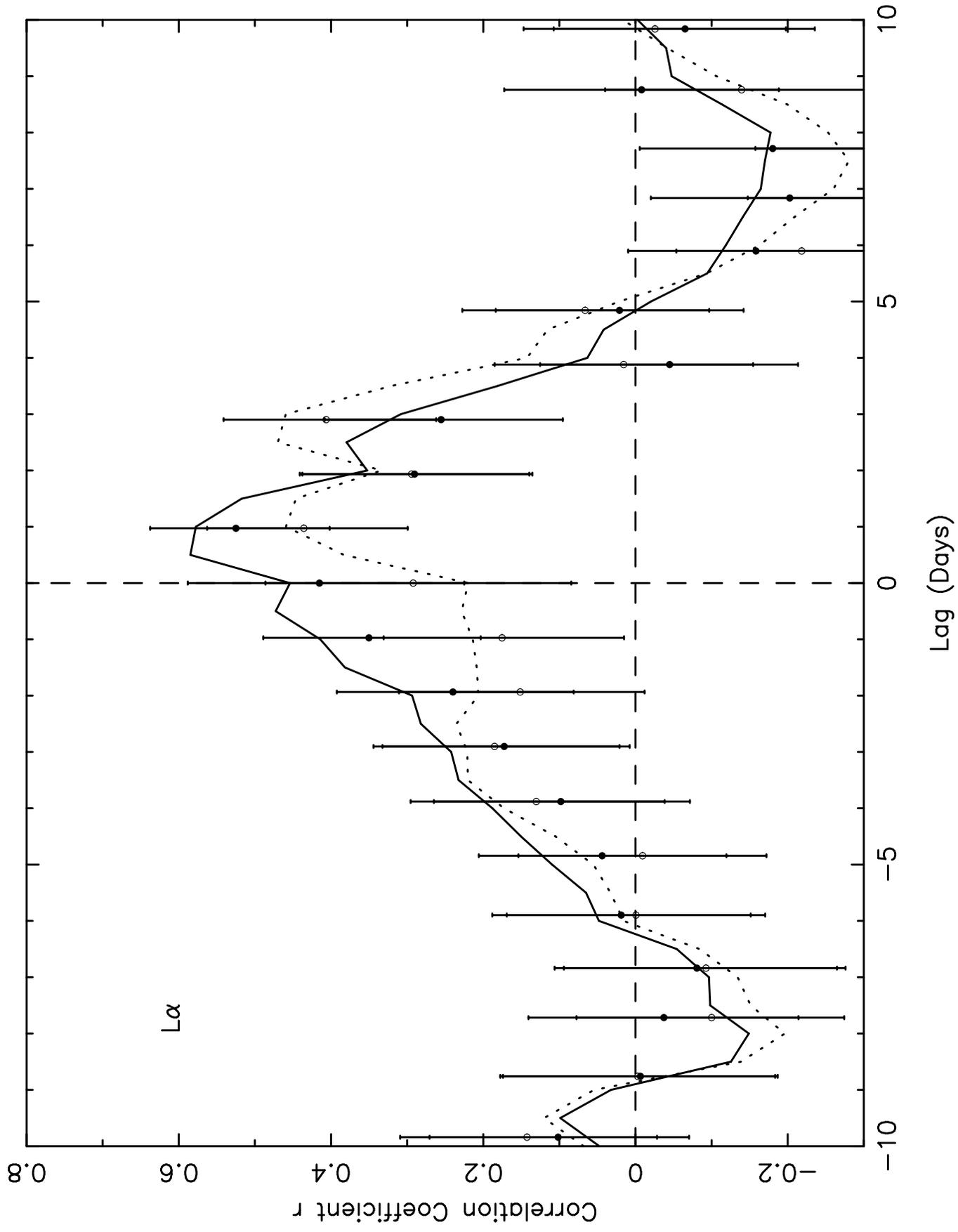
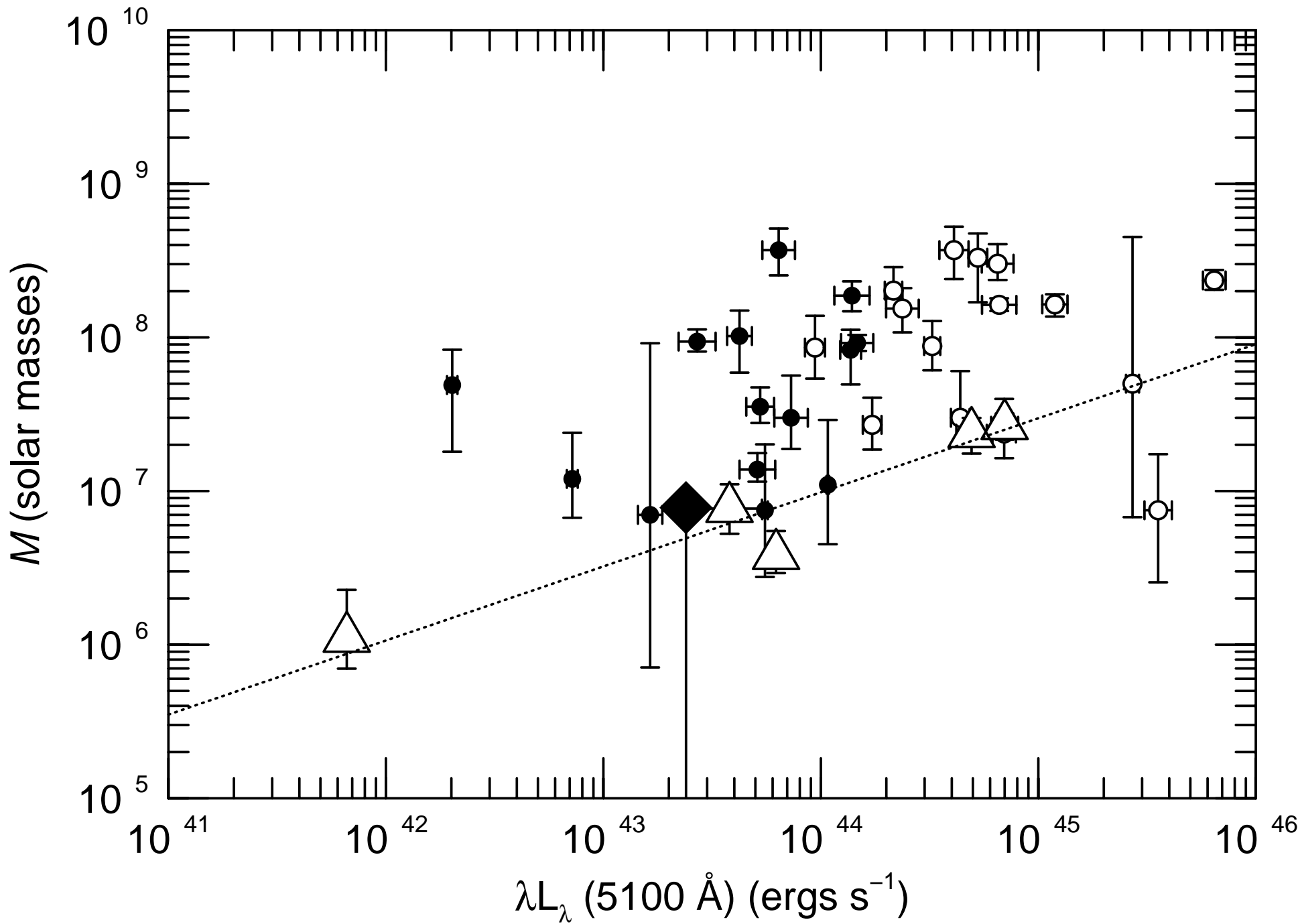
\end{document}